\documentclass[a4paper,fleqn,usenatbib]{mnras}

\usepackage{graphicx}	
\graphicspath{{./figures/}}

\usepackage{amsmath}	
\usepackage{amssymb}	

\usepackage{units}

\newcommand{\nuclei}[2]{\ensuremath{\mathrm{^{#1}#2}}}

\newcommand{\carbon}[1][12]{\nuclei{#1}{C}}
\newcommand{\nitrogen}[1][14]{\nuclei{#1}{N}}
\newcommand{\oxygen}[1][16]{\nuclei{#1}{O}}

\newcommand{\neon}[1][20]{\nuclei{#1}{Ne}}

\newcommand{\magnesium}[1][24]{\nuclei{#1}{Mg}}

\newcommand{\silicon}[1][28]{\nuclei{#1}{Si}}

\newcommand{\sulfur}[1][32]{\nuclei{#1}{S}}

\newcommand{\iron}[1][56]{\nuclei{#1}{Fe}}

\newcommand{\nickel}[1][58]{\nuclei{#1}{Ni}}

\newcommand{\code}[1]{\textsc{#1}}
\newcommand{\mesa}{\code{MESA}}
\newcommand{\MESA}{\mesa}

\newcommand{\kB}{\ensuremath{k_\mathrm{B}}} 

\newcommand{\rhoc}{\ensuremath{\rho_{\mathrm{c}}}} 
\newcommand{\gcc}{\ensuremath{\mathrm{g\,cm^{-3}}}} 

\newcommand{\Mstar}{\ensuremath{\mathrm{M}_{\star}}} 

\newcommand{\Msun}{\ensuremath{\mathrm{M}_{\sun}}} 
\newcommand{\Lsun}{\ensuremath{\mathrm{L}_{\sun}}} 

\newcommand{\Mdot}{\ensuremath{\dot{M}}} 
\newcommand{\Msunyr}{\ensuremath{\rm \Msun\,yr^{-1}}} 

\newcommand{\EF}{\ensuremath{E_\mathrm{F}}} 

\newcommand{\cp}{\ensuremath{c_{\mathrm{p}}}} 

\newcommand{\logT}{\ensuremath{\log(T/\mathrm{K})}} 
\newcommand{\logRho}{\ensuremath{\log(\rho/\gcc)}} 
\newcommand{\logRhoc}{\ensuremath{\log(\rhoc/\gcc)}} 

\newcommand{\Teff}{\ensuremath{T_{{\rm eff}}}} 
\newcommand{\logTeff}{\ensuremath{\log(\Teff/\mathrm{K})}} 

\newcommand{\rph}{\ensuremath{r_{{\rm ph}}}} 

\newcommand{\revision}{Revision 4521f0b29de258a994ca4c3aa9fca7741ed59265}

\title[Super-Chandrasekhar WD Merger Remnants]{The Evolution and Fate of Super-Chandrasekhar Mass White Dwarf Merger Remnants}

\author[Schwab, Quataert \& Kasen]{
Josiah Schwab$^{1,2}$\thanks{E-mail: jwschwab@berkeley.edu (JS)},
Eliot Quataert$^{1,2}$,
Daniel Kasen$^{1,2,3}$
\\
$^{1}${Physics Department, University of California,
     Berkeley, CA 94720, USA}  \\
$^{2}${Astronomy Department and Theoretical Astrophysics
     Center, University of California, Berkeley, CA 94720, USA} \\
$^{3}${Nuclear Science Division, 
      Lawrence Berkeley National Laboratory, Berkeley, CA 94720, USA} \\ 
}

\date{Accepted 2016 September 04. Received 2016 September 03; in original form 2016 June 06\\ \revision}

\pubyear{2016}

\begin{document}
\label{firstpage}
\pagerange{\pageref{firstpage}--\pageref{lastpage}}
\maketitle

\begin{abstract}
  We present stellar evolution calculations of the remnant of the
  merger of two carbon-oxygen white dwarfs (CO WDs).  We focus on
  cases that have a total mass in excess of the Chandrasekhar mass.
  After the merger, the remnant manifests as an
  \mbox{$L \sim \unit[3\times10^4]{\Lsun}$} source for
  $\sim \unit[10^4]{yr}$.  A dusty wind may develop, leading these
  sources to be self-obscured and to appear similar to extreme AGB
  stars.  Roughly $\sim 10$ such objects should exist in the Milky Way
  and M31 at any time.  As found in previous work, off-center carbon
  fusion is ignited within the merger remnant and propagates inward
  via a carbon flame, converting the WD to an oxygen-neon (ONe)
  composition.  By following the evolution for longer than previous
  calculations, we demonstrate that after carbon-burning reaches the
  center, neutrino-cooled Kelvin-Helmholtz contraction leads to
  off-center neon ignition in remnants with masses
  $\ge \unit[1.35]{\Msun}$.  The resulting neon-oxygen flame converts the core
  to a silicon WD.  Thus, super-Chandrasekhar WD merger remnants do
  not undergo electron-capture induced collapse as traditionally
  assumed.  Instead, if the remnant mass remains above the
  Chandrasekhar mass, we expect that it will form a low-mass iron core
  and collapse to form a neutron star.  Remnants that lose sufficient
  mass will end up as massive, isolated ONe or Si WDs.
\end{abstract}





\section{Introduction}
\label{sec:intro}

In double WD systems with sufficiently short initial orbital periods,
angular momentum losses from gravitational wave radiation shrink the
orbit and can lead to a merger of the WDs.  The outcome of such a
merger is strongly dependent on the mass of the individual WDs and
their mass ratio.  For recent summaries of the many possible outcomes,
see fig.~1 in \citet{Dan14} or fig.~3 in \citet{Shen15}.

\citet{Iben84a} and \citet{Webbink84} proposed that mergers of two
carbon-oxygen (CO) WDs whose total mass was in excess of the
Chandrasekhar mass would lead to the central ignition of carbon fusion
and hence to a Type Ia supernova.  It was quickly pointed out by
\citet{Saio85} and by \citet{Nomoto85} that the rapid mass transfer in
such an event would lead to the off-center ignition of carbon,
triggering the quiescent conversion of the remnant to an oxygen-neon
(ONe) composition.  Standard models conclude that subsequently,
electron-capture reactions cause the ONe core to collapse and form a
neutron star \citep{Miyaji80, Schwab15}.

\citet{Saio85} and \citet{Nomoto85} approximated the
effects of the WD merger as the accretion of material on to the more
massive WD.  They studied constant accretion rates
$\Mdot \la \unit[10^{-5}]{\Msunyr}$, motivated by the assumption that
the material is accreting from a disc at a rate bounded by the
Eddington limit.  Since this early work, smoothed-particle
hydrodynamics (SPH) simulations \citep[e.g.,][]{Benz90, Dan11} of
double WD systems have been used to investigate the dynamics of the
merger.  Schematically, the primary (more massive) WD remains
relatively undisturbed and the secondary (less massive) WD is tidally
disrupted.  Some material from the disrupted WD is shock-heated,
forming a thermally-supported layer at the surface of the primary WD;
the rest of the material is rotationally-supported, forming a thick
disc at larger radii.\footnote{Exceptions to this picture include WD
  collisions \citep[e.g.,][]{Raskin09} or cases where the mass ratio is
  nearly unity \citep[e.g.,][]{vanKerkwijk10}.}

\citet{Yoon07} took an important step forward by constructing 1D
models that mimicked the results of SPH simulations of WD mergers.
They then used a hydrodynamic stellar evolution code to evolve these
models, allowing them to follow the secular evolution of the remnants.
However, \citet{Shen12} and \citet{Schwab12} studied the post-merger
evolution of these systems in more detail and showed that the
transport of angular momentum via magnetic stresses occurs on a
time-scale ($\sim\unit[10^{4}]{s}$) far shorter than the time-scale on
which the remnant cools ($\sim\unit[10^{4}]{yr}$).  Material initially
in the disc quickly became thermally supported and spherical.  In the
models of \citet{Yoon07}, who neglected the effects of magnetic
fields, the time-scale for angular momentum redistribution was often
comparable to or longer than the cooling time. The disc was assumed to
be long-lived and modeled as a low accretion rate on to the central
remnant.  One of the key conclusions of \citet{Shen12} and
\citet{Schwab12} was that the rapid viscous evolution should be taken
into account before exploring the long-term thermal evolution of the
merger remnant.
 

In this work, we follow the long-term evolution of the merger of two
CO WDs.  The initial conditions for these calculations are generated
self-consistently, beginning from SPH simulations of the dynamical
merger of the two WDs \citep{Dan11, Raskin14} and explicitly modeling
the subsequent phase of viscous evolution as in \citet{Schwab12}.  We
then use the \MESA\ stellar evolution code
\citep{Paxton11,Paxton13,Paxton15} to follow the evolution of these
merger remnants over thermal and nuclear time-scales.


In Section~\ref{sec:setup} we discuss our initial models along with
the key options and input physics that enable our \MESA\ calculations.
In Section~\ref{sec:cflame} we discuss the ignition and propagation of
a carbon flame.  In Section~\ref{sec:kh} we discuss the ignition of a
neon-oxygen flame and then in Section~\ref{sec:subsequent-evolution} outline
how subsequent evolution may lead to the formation of a neutron
star. In Section~\ref{sec:observ-prop} we discuss the observational
properties of the remnant.  In Section~\ref{sec:conclusions} we
conclude, presenting a schematic overview of our results and
suggesting avenues for future work.  Fig.~\ref{fig:flowchart}
summarizes the possible end states of super-Chandrasekhar WD mergers.

\section{Setup of \MESA\ Calculations}
\label{sec:setup}

\subsection{Initial Models}
\label{sec:initial-models}

As discussed in the introduction, SPH simulations of WD mergers show
that at the end of the dynamical phase of the merger, the primary
white dwarf remains mostly undisturbed.  It is now surrounded by the
tidally disrupted secondary, which includes a significant amount of
material in a rotationally-supported disc.  In \citet{Schwab12}, we
took the output of SPH merger calculations, mapped them into a
grid-based hydrodynamics code \citep[ZEUS-MP2;][]{Hayes06},
and followed the evolution of these
remnants under the action of an $\alpha$-viscosity.  We found that the
viscous stresses transformed the disc material into a spherical,
thermally-supported envelope on a time-scale of hours.

In this work, we focus on two super-Chandrasekhar WD merger remnants,
one with a total mass of approximately $1.5\,\Msun$ (model M15) and
another with total mass of approximately $1.6\,\Msun$ (model M16).
Both remnants are the result of a merger with mass ratio $q = 2/3$ and
have a composition of 50 per cent carbon and 50 per cent oxygen by
mass.  Table~\ref{tab:models} contains a summary of these models,
including references to the papers in which the simulations of the
dynamical and viscous phases were first reported.  

\begin{table}
  \centering

  \begin{tabular}{llllll}
    \hline
    ID & SPH Ref. & $M_2$ & $M_1$  & Viscous Ref. & $M_\mathrm{tot}$ \\
    \hline
    M15 & Dan11 & 0.60 & 0.90 & Schwab12 & 1.486 \\
    M16 & Raskin14 & 0.64 & 0.96 & Raskin14 & 1.586 \\
    \hline
  \end{tabular}

  \caption{A summary of the two merger systems studied in this
    work. The ID reflects the total mass of the system. ``SPH ref.''
    indicates the primary reference containing the details of the SPH
    calculation of the merger (Dan11: \citealt{Dan11}, Raskin14:
    \citealt{Raskin14}).  $M_2$ is the mass (in $\Msun$) of the
    secondary, the less massive of the two WDs; $M_1$ is the mass of
    the more massive primary. ``Viscous ref.'' refers to the primary
    reference containing the details of the subsequent viscous
    evolution (Schwab12: \citealt{Schwab12}, Raskin14:
    \citealt{Raskin14}). $M_\mathrm{tot}$ is the total mass of the
    bound remnant (in \Msun) at the end of the viscous phase simulation, and
    hence the initial mass of the \MESA\ model.}
  \label{tab:models}
\end{table}

\begin{figure}
  \centering
  \includegraphics[width=\columnwidth]{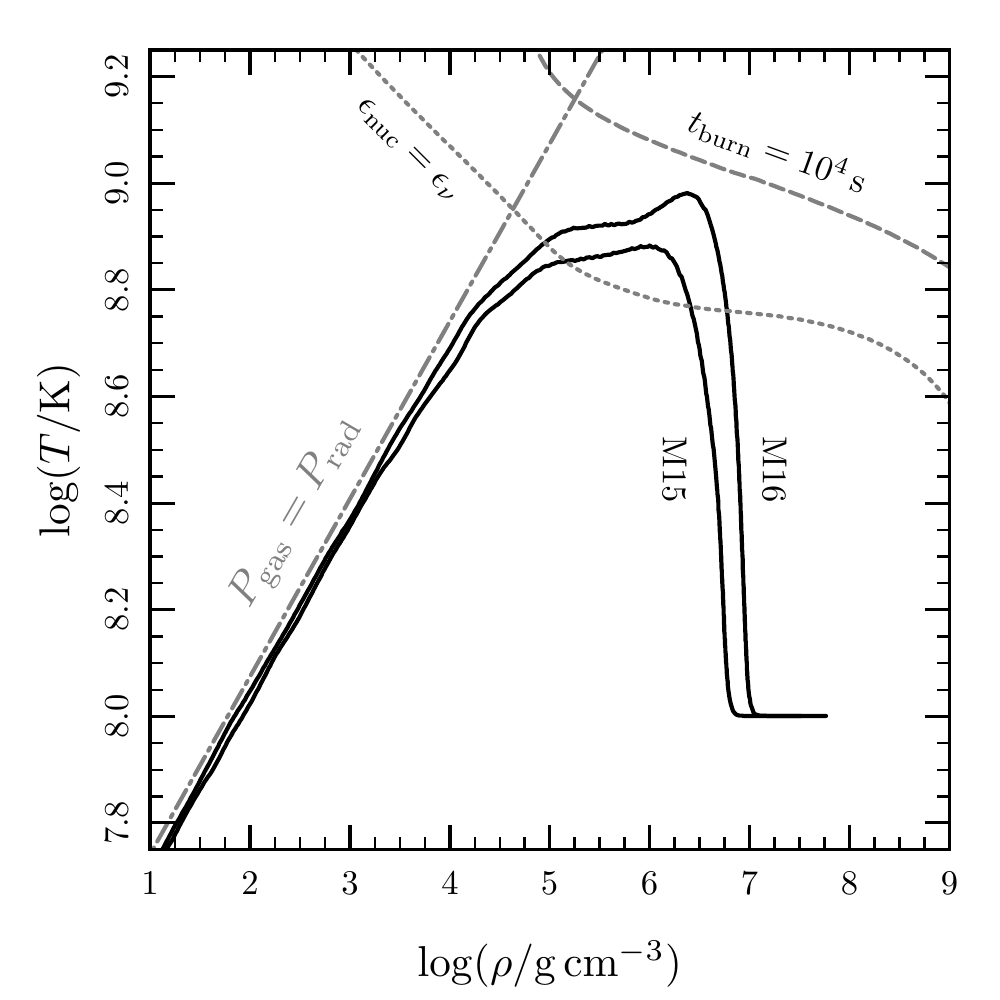}
  \caption{The initial density-temperature profiles of our \MESA\
    models M15 and M16, taken from the end state of simulations of the
    viscous phase of WD merger remnants (see
    Appendix~\ref{sec:mapping-models}).  The dotted line shows where
    the energy release from nuclear burning is equal to the thermal
    neutrino losses for an equal (by mass) mixture of carbon and
    oxygen. The dashed line shows where the burning time-scale
    $t_\mathrm{burn} = \unit[10^4]{s}$.  The dash-dot line indicates
    where gas and radiation pressure are equal.  Both models reached
    carbon ignition conditions during the viscous evolution, but the
    duration of the viscous phase was not long enough to allow
    significant nuclear burning to occur.  Since the structure of the
    $\unit[1.6]{\Msun}$ model is so similar to that of the
    $\unit[1.5]{\Msun}$ model, its evolution and outcomes are similar
    (aside from the possible effects of mass loss; see
    Section~\ref{sec:kh}).}
  \label{fig:initial-models}
\end{figure}

The initial conditions for the thermal evolution calculations in this
work are the end states of our viscous evolution calculations.  In
Appendix~\ref{sec:mapping-models}, we discuss the details of how we
transfer the output of our hydrodynamic simulations into \MESA.  In
Fig.~\ref{fig:initial-models}, we show the density-temperature
profiles of our initial \MESA\ models.  We define the burning time as
$t_{\mathrm{burn}} = \cp T / \epsilon_\mathrm{nuc}$, where $\cp$ is
the specific heat at constant pressure, $T$ is the temperature and
$\epsilon_{\mathrm{nuc}}$ is the specific energy generation rate from
nuclear reactions.  At the temperature peaks of both models,
$t_{\mathrm{burn}} \ga \unit[10^4]{s}$, the approximate duration of the
viscous evolution.  This demonstrates that substantial nuclear energy was
not liberated during the viscous phase.

\subsection{\MESA\ Version and Important Options}
\label{sec:options}

The calculations performed in this paper use \MESA\ version r6596
(released 2014-06-08).  As required by the \MESA\ manifesto, the
inlists necessary to reproduce our calculations will be posted on
\url{http://mesastar.org}.

We use \texttt{approx21.net}, a 21-isotope $\alpha$-chain nuclear
network.  In order to avoid spurious flashes, it is important to
resolve the burning fronts in the convectively-bounded deflagrations
that develop in our models \citep[e.g.,][]{Saio98}.  We use the
options selected by \citet{Farmer15} in their study of carbon flames
in super-AGB stars.  In particular, we use the controls
\begin{verbatim}
    mesh_dlog_burn_c_dlogP_extra = 0.10
    mesh_dlog_cc_dlogP_extra = 0.10
    mesh_dlog_co_dlogP_extra = 0.10
\end{verbatim}
which add additional spatial resolution in regions where carbon
burning is occurring.  This ensures that that flame is well-resolved.
We use analogous controls to ensure that the neon-oxygen-burning flames in
our models are also well-resolved.

\subsection{Input Physics}
\label{sec:opacities}

The initial composition of the model is pure carbon and oxygen.  We
use the OPAL radiative opacities for carbon and oxygen-rich mixtures
\citep{Iglesias93, Iglesias96}.  These are referred to as OPAL ``Type
2'' tables in \MESA.  We select a base metallicity using the control
$\mathtt{Zbase} = 0.02$.  The lower temperature boundary of these
tabulations is $\logT = 3.75$.

As we show in Section~\ref{sec:observ-prop}, when the outer layers of
the remnant expand, they reach temperatures below $\logT = 3.75$.
\MESA\ does not provide low-temperature opacities that include
separate carbon and oxygen enhancements.  As a result, \MESA\ is
usually forced to fall back to opacity tabulations which assume a
different composition.  Thus, when blending between the OPAL tables
and any of the included low-temperature tables, there are dramatic and
unphysical changes in opacity at the location of the blend.

In order to ensure that the composition assumed by the opacities
approximately matches the composition of the model, we generate a new
opacity table.  These calculations and their results are briefly
described in Appendix~\ref{sec:kasen-opacities}.  For the opacities,
we consider only a single composition: $dX_\mathrm{C} = 0.49$,
$dX_\mathrm{O} = 0.49$, $Z = 0.02$, where the relative metal
abundances are drawn from \citet{Grevesse98}.\footnote{Here,
  $dX_\mathrm{C}$ and $ dX_\mathrm{O}$ indicate the mass fraction
  enhancements above the (relatively small) amount of carbon and
  oxygen already included in $Z$.  Also, the stellar evolution
  preceding the formation of the individual WDs would have modified
  some of the abundances contained in the standard solar Z (e.g.,
  \nitrogen[14] would have been processed to \neon[22] during helium
  burning) but we neglect those changes.}  We do not consider the
effects of molecular opacities in these calculations, putting a rough
lower limit on their validity of $\logT \ga 3.5$.  Using these tables,
our models obey $\logTeff > 3.6$ and so do not violate this
assumption.  We discuss the role of molecule and dust formation in
these objects in Section~\ref{sec:observ-prop}.




\section{Carbon Flame}
\label{sec:cflame}

\begin{figure}
  \centering
  \includegraphics[width=\columnwidth]{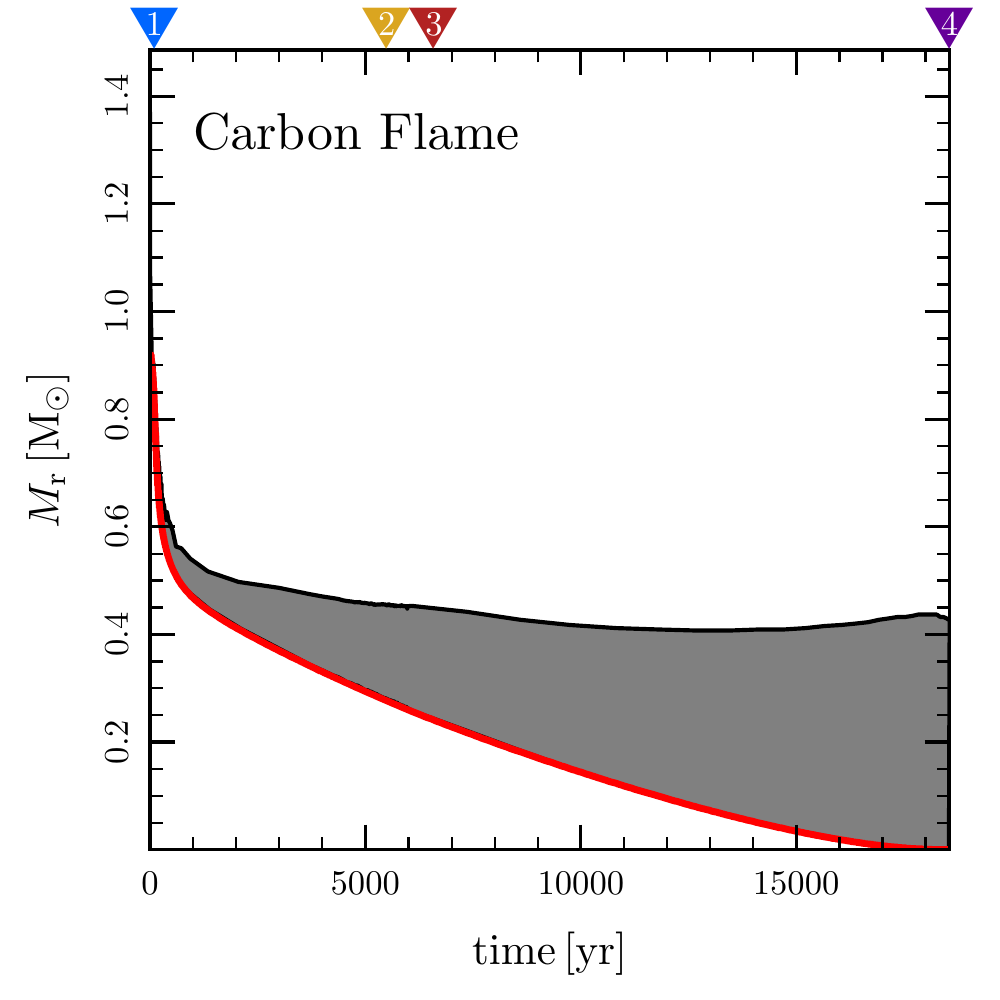}
  \caption{The propagation of the carbon flame in model M15 with no
    convective boundary mixing.  The x-axis shows time, as measured
    from the beginning of the \MESA\ calculation; effectively this is
    the time since merger.  The y-axis shows the Lagrangian mass
    coordinate.  The extent of the convective region associated with
    the flame is shaded.  The location of maximum nuclear energy
    release, a proxy for the location of the flame, is indicated by
    the thick red line at the bottom of this region.  After
    approximately 20 kyr, the flame reaches the center.  The numbered
    triangles at the top of the plot indicate times in the evolution
    that will be referenced in Figs.~\ref{fig:center-M15} and
    \ref{fig:hr-M15}.}
  \label{fig:flame-M15}
\end{figure}

In each of the initial models shown in Fig.~\ref{fig:initial-models},
the rate of energy release from nuclear reactions exceeds the rate of
energy loss from thermal neutrino cooling at the initial temperature peak.  As
described in \citet{Schwab12}, this is because heating by viscous
dissipation leads to off-center ignition of self-sustaining carbon
fusion within hours of the merger.\footnote{This is in contrast to
  calculations that model the merger as Eddington-limited accretion,
  in which carbon ignition does not occur for $\ga \unit[10^{4}]{yr}$,
  until sufficient material has accreted to adiabatically compress the
  base of the shell to higher temperatures.  In a lower mass merger,
  say $M_{\mathrm{tot}} \approx \unit[1.4]{\Msun}$, we find that
  carbon fusion does not ignite during the viscous evolution.  In this
  case, there is a similar time delay, as carbon ignition must wait
  for the cooling envelope to compress material at its base.  An
  example of such evolution is shown in \citet{Shen12}.}  When we
begin our \MESA\ calculation, there is thus immediately off-center
carbon burning in the remnant.  The energy release in this
carbon-burning shell quickly leads to the formation of a convection
zone.  Heat from the burning region is conducted into the degenerate
interior, giving rise to a deflagration wave that begins propagate
towards the center of the remnant.  We refer to this deflagration as
the ``carbon flame''.  Fig.~\ref{fig:flame-M15} shows the evolution of
this flame in our \MESA\ calculations of model M15.  After an initial
transient phase with a duration of $\la 100$ years ($t=0$ to time 1),
the deflagration forms and propagates to the center over
$\approx \unit[20]{kyr}$ (time 1 to time 4).  Because the structure
of the $\unit[1.6]{\Msun}$ model is so similar to that of the
$\unit[1.5]{\Msun}$ model (see Fig.~\ref{fig:initial-models}) its
evolution and outcomes are similar; for simplicity, we will primarily
discuss model M15.

It is important to note that the carbon flame, while off-center, is
still deep in the interior.  The convective zone outside the burning
region satisfies a ``balanced power'' condition, where the total
luminosity of thermal neutrino emission from the zone is approximately
equal to the rate of energy release from fusion at its base
\citep{Timmes94}.  This neutrino-cooled convective zone has a radial
extent of order the pressure scale height, and its upper boundary
is sufficiently deep that the time-scale for radiative diffusion to
transport the energy to the surface is longer than the evolutionary
time-scale of the remnant.  Thus, while key to the final fate of the
remnant, the energy release of the carbon flame is not coupled to the
surface, and does not power the luminosity of the remnant.  Instead,
the behavior of the surface layers---which we discuss in more detail
in Section~\ref{sec:observ-prop}---is driven by the thermal energy
generated during the merger.

\begin{figure}
  \centering
  \includegraphics[width=\columnwidth]{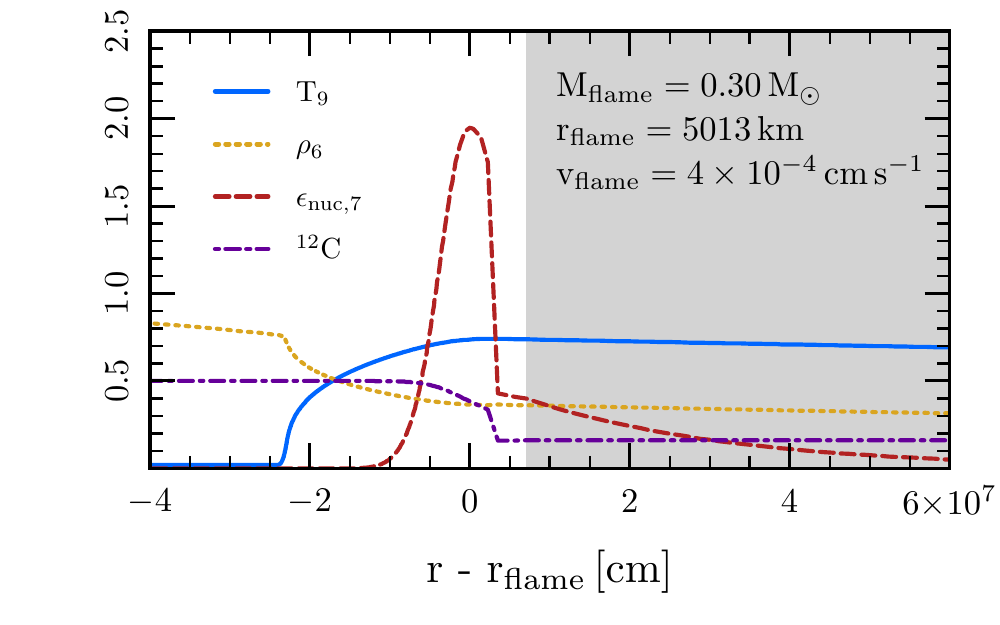}
  \caption{Structure of a carbon-burning flame.  The shaded grey
    region marks the convection zone. The temperature
    $(T/\unit[10^9]{K})$, density $(\rho/\unit[10^6]{g\,cm^{-3}})$,
    energy generation rate
    $(\epsilon_{\mathrm{nuc}}/\unit[10^7]{erg\,s^{-1}\,g^{-1}})$, and
    \carbon[12] mass fraction are shown as function of radius.  The
    thickness of the flame is $\sim \unit[10^7]{cm}$.  This profile
    from our \MESA\ calculation is from a time slightly before time
    2, as marked in Fig.~\ref{fig:flame-M15}.}
  \label{fig:cflame_structure}
\end{figure}

A snapshot of the carbon flame structure is shown in
Fig.~\ref{fig:cflame_structure}.  At this time, the flame is at a
density $\rho \approx \unit[4\times10^5]{g\,cm^{-3}}$ and temperature
$T \approx \unit[7\times10^8]{K}$, with a carbon mass fraction
$X_{\rm C} \approx 0.5$.  At these conditions, the flame width is
$\approx \unit[2\times10^7]{cm}$ and the flame speed is
$\approx \unit[4\times10^{-4}]{cm\,s^{-1}}$.  While this exact
density, temperature, and composition are not present in
\citet{Timmes94}, the flame speed and thickness we find are
consistent with their tabulated results.


\citet{Denissenkov13b} suggested that efficient mixing at the
convective boundary can quench inwardly propagating carbon flames in
super-AGB stars.  If a similar phenomenon were to occur here, the
death of the carbon flame would lead to qualitatively different
results, as it would create a ``hybrid'' WD with a CO core and an ONe
mantle.  We include this possibility in the flow chart presented in
Section~\ref{sec:conclusions} (Fig.~\ref{fig:flowchart}).  However, \citet{Lecoanet16b} find that
convective plumes fail to induce sufficient mixing to lead to flame
disruption and conclude that these ``hybrid'' WDs are not a typical
product of stellar evolution.

\section{Kelvin-Helmholtz Contraction and Neon Ignition}
\label{sec:kh}

\begin{figure}
  \centering
  \includegraphics[width=\columnwidth]{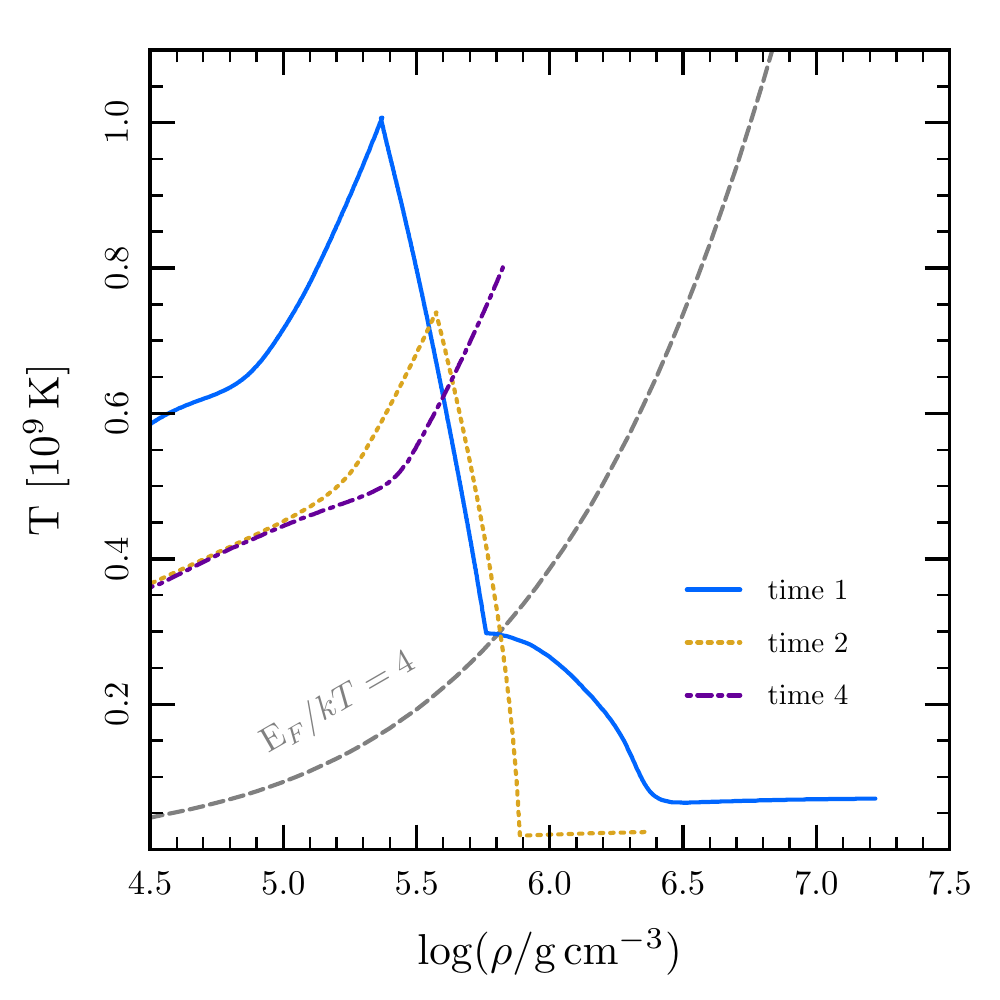}
  \caption{The evolution of the core of model M15 in
    temperature-density space as the carbon flame propagates inward.
    Each line corresponds to one of the times indicated in
    Fig.~\ref{fig:flame-M15}; time 3 is omitted because it appears
    extremely similar to time 2.  The location of the flame
    corresponds to the location of maximum temperature in each line.
    The dashed grey line marks the degeneracy condition (above/left is
    non-degenerate; below/right is degenerate).  Early in the
    evolution (time 1) the center has $\logRho > 7$ and is degenerate.  As the flame
    propagates inward, it lifts the degeneracy of additional material,
    and the central density decreases (time 2).  Once the flame
    reaches the center (time 4), it has lifted the degeneracy
    throughout the star.}
  \label{fig:center-M15}
\end{figure}

Once the carbon flame reaches the center, it has lifted the degeneracy of
the material throughout the star.  This is illustrated in Fig.~\ref{fig:center-M15}, where
the dash-dot line (labeled time 4) shows the temperature-density profile of the model
at the time of central carbon exhaustion.  The newly non-degenerate
core will now Kelvin-Helmholtz (KH) contract.  The core is
sufficiently hot and dense that it cools through thermal neutrino
losses.  As a result, it will develop an off-center temperature peak.
We note that the evolution of the central density and temperature is
similar to that seen in models of intermediate mass stars $(\approx \unit[8-10]{\Msun})$, as these objects
also develop degenerate cores with similar masses
\citep[e.g.,][]{Jones13,Jones14}.

As discussed by \citet{Nomoto84a}, there is a critical core mass for
off-center neon ignition.  Nomoto demonstrated this by means of a
simple calculation in which stellar models of pure neon were
constructed and allowed to KH contract.  Neon ignition occurred only
in models with a mass above $1.37 \Msun$.  We repeat this
calculation and find a slightly lower critical mass of $1.35 \Msun$ for
pure $\neon$ models (see Appendix~\ref{sec:critical-mass}).  We use
the results of these pure neon models to guide our interpretation of
the central evolution of the WD merger remnants.

\begin{figure}
  \centering
  \includegraphics[width=\columnwidth]{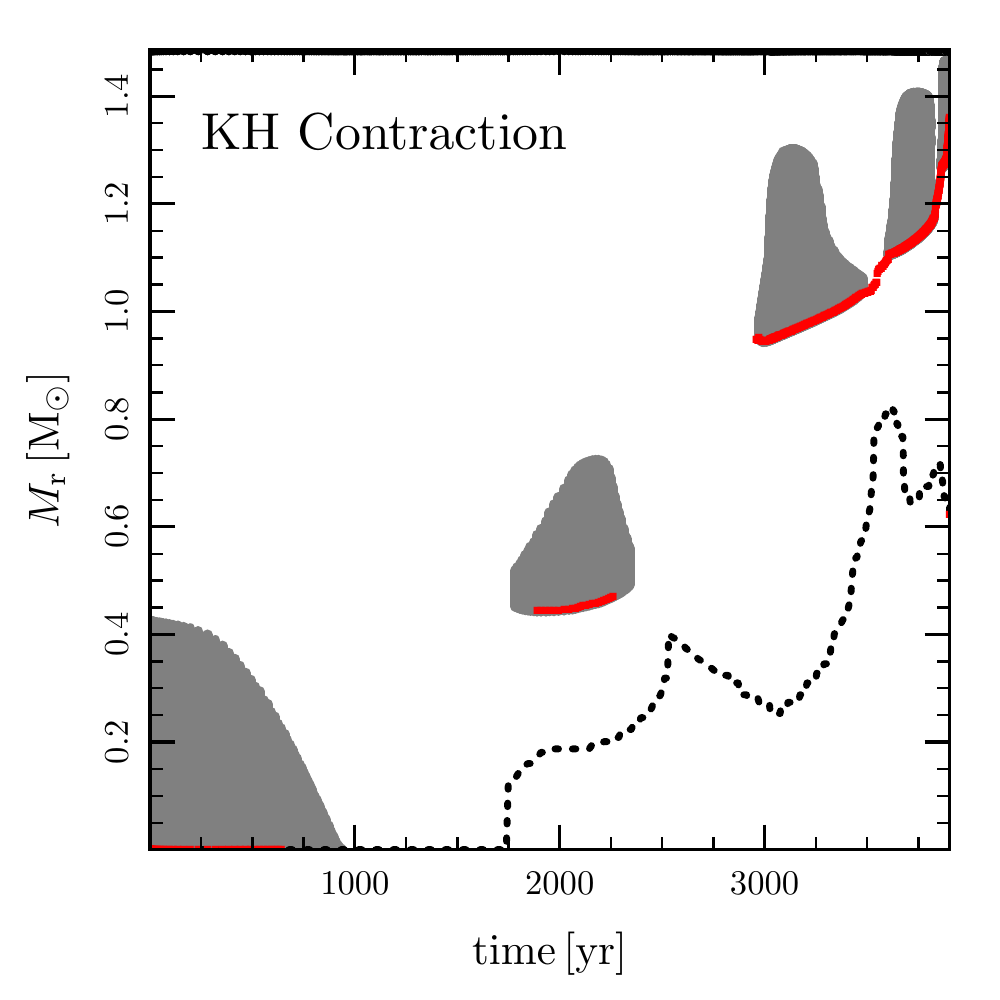}
  \caption{A Kippenhahn diagram of model M15 from the time the carbon
    flame reaches the center until off-center neon ignition.  The x-axis shows
    time, as measured from the time when the carbon flame reached the
    center.  The y-axis shows the Lagrangian mass
    coordinate. Convective regions are shaded grey and the locations
    of carbon burning (with
    $\epsilon_{\rm nuc} > \unit[10^7]{erg\,s^{-1}\,g^{-1}}$) are
    marked in red.  The dotted black line indicates the location of the
    local maximum in the temperature profile that is closest to the center.}
  \label{fig:kh-kipp-M15}
\end{figure}

Fig.~\ref{fig:kh-kipp-M15} shows a Kippenhahn diagram of model M15
from the time the carbon flame reaches the center until neon ignition.
Initially carbon is being burned in the core, but over the first
$\approx \unit[1]{kyr}$, the central convection zone shrinks and vanishes as
the carbon in the center is exhausted.  As the star KH contracts, a
series of off-center carbon flashes occur.  Additionally, an
off-center temperature peak develops.  Its mass coordinate is
indicated by the black dotted line.

\begin{figure}
  \centering
  \includegraphics[width=\columnwidth]{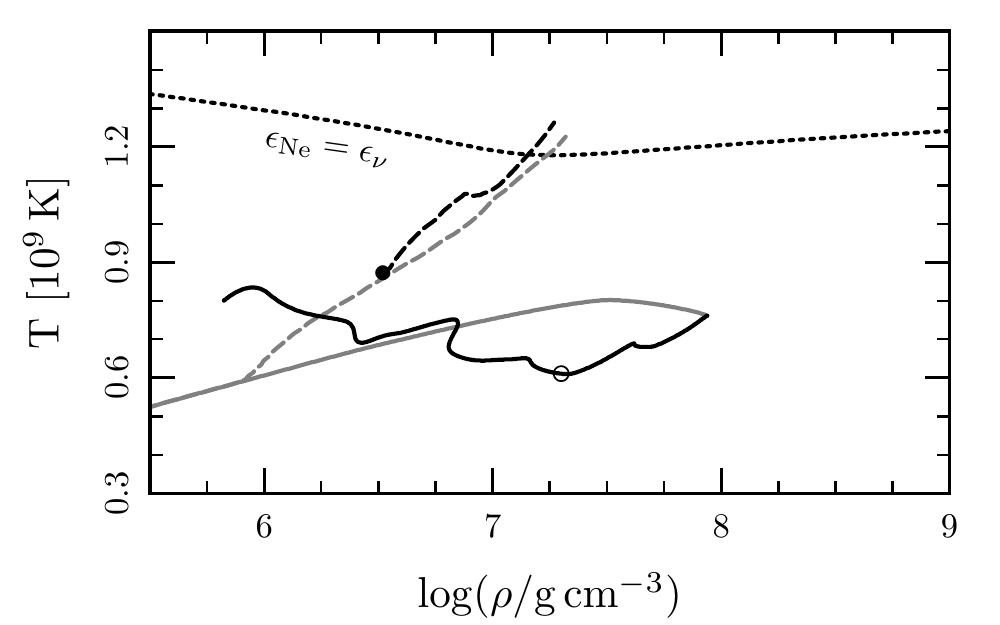}
  \caption{The evolution of temperature and density at the center of
    model M15 (solid black line) and at the temperature peak (dashed
    black line) during the KH contraction shown in
    Fig.~\ref{fig:kh-kipp-M15}.  The total mass of the remnant is
    $\approx 1.5\,\Msun$.  For visual clarity, the line for the
    temperature peak is shown only after the central density exceeds
    $\log(\rho/\gcc) \approx 7.3$; the first plotted point of the
    dashed (temperature peak) line is marked by a solid black circle
    and the contemporaneous point on the solid (center) line is marked
    by an open circle.  The grey lines show the evolution of a
    contracting $1.385\,\Msun$ pure neon model; the solid grey line
    shows conditions at the center and the dashed grey line shows
    conditions at the temperature peak, which due to neutrino cooling,
    develops off-center.  The black dotted line shows approximately
    where the energy release from neon burning is equal to the energy
    loss rate from thermal neutrinos.  Off-center neon ignition occurs
    in this model.}
  \label{fig:kh-M15-center}
\end{figure}

Fig.~\ref{fig:kh-M15-center} shows the evolution of the temperature
and density at both the center (solid black line) and the off-center
temperature peak (dashed black line) in model M15 during the KH
contraction phase.  The ``wiggles'' in the evolution of the center are
manifestations of changes to the stellar structure due to the
off-center carbon flashes shown in Fig.~\ref{fig:kh-kipp-M15}.  The
grey lines show the analogous temperature and
density evolution of a $1.385 \Msun$ pure neon model.  The qualitative
agreement between the two models is good.

The agreement between our full remnant models and our
simple pure neon calculations demonstrates that the off-center neon
ignition is a simple consequence of the mass of the remnant.  Once
ignited, the neon burning will propagate to the center in a manner
similar to the carbon burning, converting the object to silicon-group
elements.  (See Section~\ref{sec:subsequent-evolution} for more
discussion of this process and the subsequent evolution.)  Because the
critical mass for neon ignition $(1.35 \Msun)$ is less than the
critical mass needed to trigger the collapse of an ONe core
\citep[$1.38\Msun$;][]{Schwab15}, we conclude that is is
difficult to produce an ONe core with a sufficient mass to undergo AIC
in a WD merger.

Off-center neon ignition does not appear to have been observed in
previous studies of CO WD merger remnants.  The salient difference
between this work and previous work appears to be that we evolved the
remnants for longer.  \citet{Saio85} halted their calculation when the
carbon flame was at a mass coordinate of $M_r = 0.005\,\Msun$ because
it had become too computationally costly to continue.  Later work by
the same authors \citep{Saio98} allowed the flame to reach the center,
but did not continue the evolution beyond this point.  The work by
\citet{Yoon07}, which this work is most similar to, focused on avoiding
off-center carbon ignition; in cases where off-center carbon ignition
did occur, the authors did not continue their evolutionary calculations.

\begin{figure}
  \centering
  \includegraphics[width=\columnwidth]{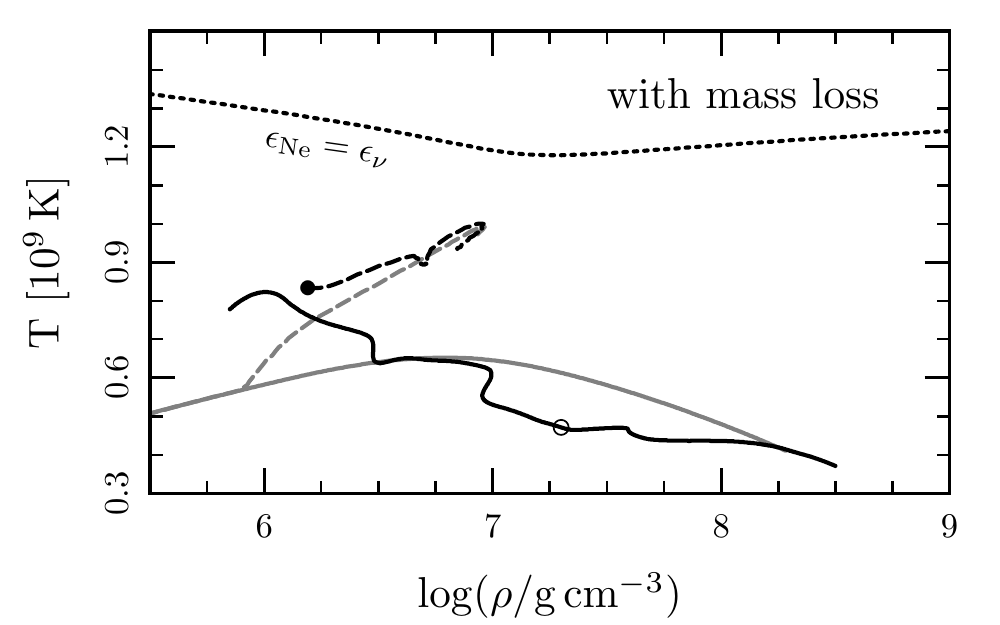}
  \caption{Same as Fig.~\ref{fig:kh-M15-center}, but for the version
    of model M15 with mass loss during the cool giant phase (see
    Section~\ref{sec:kh}).  The remnant shed
    $\approx 0.2\,\Msun$ and now has a total mass of
    $\approx 1.3\,\Msun$.  The grey lines show the evolution of a
    contracting $1.30\,\Msun$ pure neon model.  Neon ignition does not
    occur and the core instead becomes supported by electron
    degeneracy pressure.  The remnant will cool to form an ONe WD.}
  \label{fig:kh-M15-center-mass-loss}
\end{figure}


As an illustration of the effects of mass loss on the core, we run a
version of model M15 with a mass loss rates drawn from
\citet{Bloecker95a}.  That is, we use the \MESA\ options
\begin{verbatim}
    AGB_wind_scheme = 'Blocker'
    Blocker_wind_eta = 0.1d0
\end{verbatim}
The \citet{Bloecker95a} mass loss rates were motivated by atmosphere
calculations of Mira-like stars and thus they are not directly
applicable to this problem.  However, we simply want to remove some
mass when the object is cool and luminous, and so this is a suitable
heuristic.  We discuss mass loss in more detail in
Section~\ref{sec:observ-prop}.

Fig.~\ref{fig:kh-M15-center-mass-loss} shows the same quantities as
Fig.~\ref{fig:kh-M15-center}, but for a model in which the remnant
shed approximately $0.2\,\Msun$ of material during the phase while the
carbon flame was propagating to the center.  As the object contracts,
an off-center temperature peak develops, but it fails to reach
temperatures where neon burning exceeds thermal neutrino
losses.  Instead, the core becomes supported by electron degeneracy
pressure, halting the KH contraction.  As a result, the peak
temperature reaches a maximum and then begins to decrease.  The
remnant will retain its ONe composition and cool to become a massive WD.

\section{Neon-Oxygen Flame and Subsequent Evolution}
\label{sec:subsequent-evolution}

The phase beginning with off-center neon ignition is relatively
unexplored.  Although, we are unable to self-consistently evolve the
remnants all the way to their final fate (as discussed in
Section~\ref{sec:siflame}), the likely outcome for sufficiently
massive remnants is the formation of a neutron star via an iron core
collapse.  Analogous phases of evolution in single intermediate mass
stars \citep[{$\approx \unit[8-10]{\Msun}$;}][]{Jones14, Woosley15}
and in ultra-stripped binary systems \citep[i.e., ones that form
helium cores 2.5-3.5 \Msun;][]{Tauris13b, Tauris15} continue to be
active areas of research.  Future results in these areas can be
applied to the evolution of super-Chandrasekhar mass WD merger
remnants.

\subsection{Neon-Oxygen Flame}
\label{sec:neflame}

\begin{figure}
  \centering
  \includegraphics[width=\columnwidth]{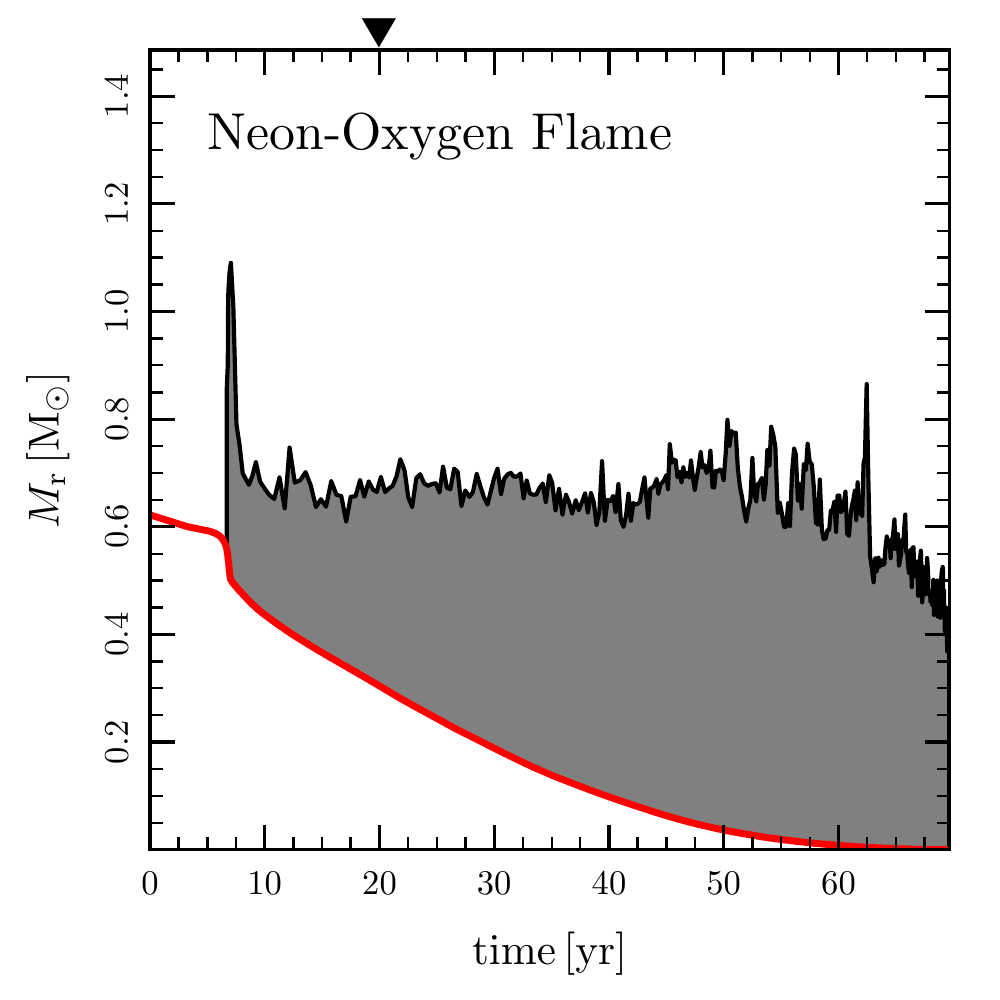}
  \caption{The propagation of the neon-oxygen flame in model M15 with no
    convective boundary mixing.  The x-axis shows time, as measured
    from the beginning of neon burning.  Note that the neon-oxygen flame
    propigates to the center $\sim100$ times more quickly than the
    carbon flame shown in Fig.~\ref{fig:flame-M15}.  The y-axis shows
    the Lagrangian mass coordinate.  The extent of the convective
    region associated with the flame is shaded. The location of the
    outer boundary of the convection zone varies significantly from
    timestep-to-timestep, so for visual clarity, this has been
    smoothed.  The location of maximum nuclear energy release, a proxy
    for the location of the flame, is indicated by the red thick line
    at the bottom of this region.  The triangle at the top of the plot
    marks the time at which the flame structure is shown in
    Fig.~\ref{fig:neflame_structure}.}
  \label{fig:neflame}
\end{figure}

In model M15, off-center neon ignition occurs at a mass coordinate
$M_r \approx 0.6\,\Msun$.  This quickly forms a convectively-bounded
neon-oxygen-burning deflagration front which begins to propagate inwards
towards the center of the star.  Fig.~\ref{fig:neflame} shows the
location of the flame and its accompanying convection zone during this
phase.  In our \MESA\ calculations, we assume that there is no mixing (i.e., convective overshoot or thermohaline mixing) beyond the convective boundary.

\begin{figure}
  \centering
  \includegraphics[width=\columnwidth]{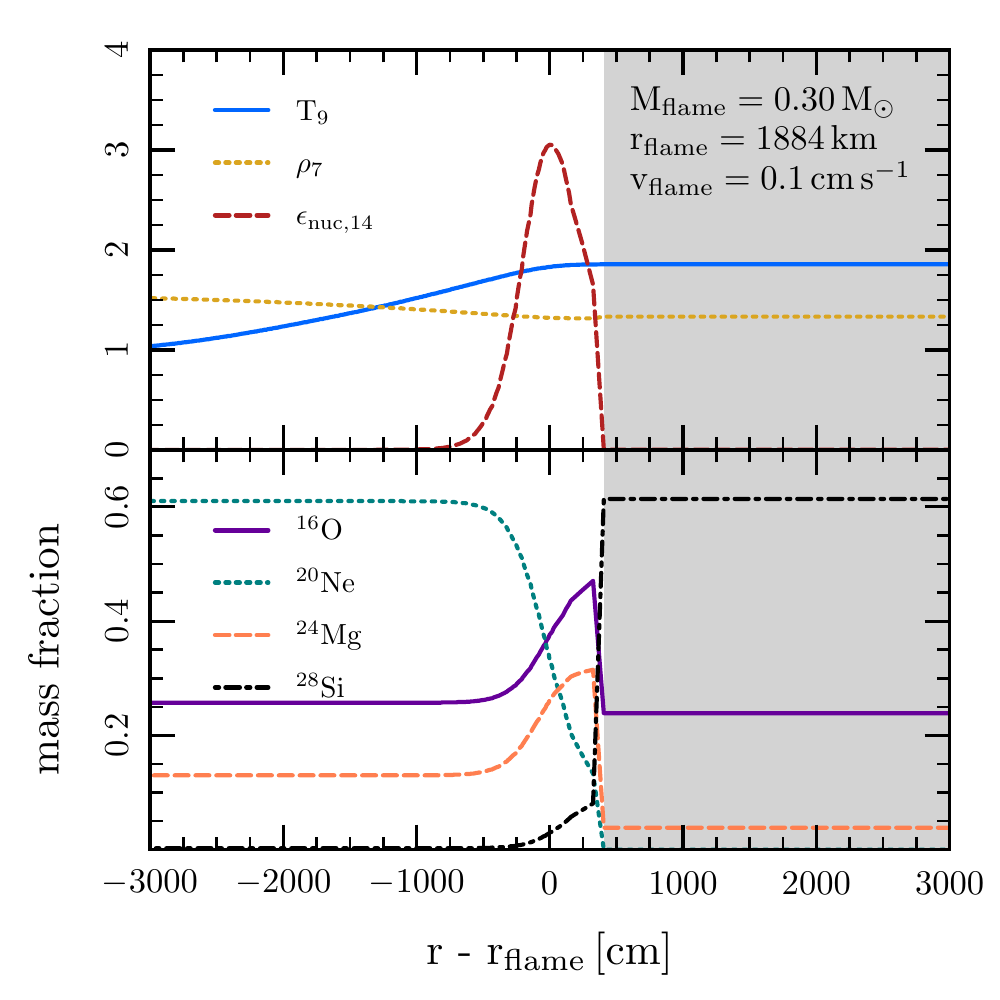}
  \caption{Structure of a neon-oxygen-burning flame.  The shaded grey
    region marks the convection zone.  The top panel shows the
    temperature $(T/\unit[10^9]{K})$, density
    $(\rho/\unit[10^7]{g\,cm^{-3}})$, and energy generation rate
    $(\epsilon_{\mathrm{nuc}}/\unit[10^{14}]{erg\,s^{-1}\,g^{-1}})$ as
    function of radius, centered around the flame.  This illustrates
    that the thickness of the flame is $\sim \unit[10^3]{cm}$, which
    makes directly resolving the propagation of the
    neon-oxygen-burning flame computationally costly.  The bottom
    panel shows the abundances of key isotopes.  Photodisintegration
    of \neon[20] produces \oxygen[16] and \magnesium[24], with
    $\alpha$-captures on \magnesium[24] producing \silicon[28].  The
    \silicon[28] (and \sulfur[32], not shown) abundance is larger in
    the convection zone than at the ``back'' of the flame because
    further $\alpha$-captures and oxygen-burning occur in the bounding
    convection zone.  This profile from our \MESA\ calculation is from
    the time marked by the black triangle in Fig.~\ref{fig:neflame}.}
  \label{fig:neflame_structure}
\end{figure}

As can be shown from the analytic estimates given in \citet{Timmes94}
and as discussed in \citet{Woosley15}, neon-oxygen-burning flames are
much faster (though still extremely subsonic) and thinner than
carbon-burning flames.  This is due to the higher temperatures and
energy generation rates associated with neon and oxygen fusion.
Neon-burning consists of photo-disintegration
$\neon[20] + \gamma \to \oxygen[16] + \alpha$ followed by
$\alpha$-captures, with the net result being the rearrangement
$2\,\neon[20] \to \oxygen[16] + \magnesium[24]$ \citep{Woosley02}.
Subsequently, additional $\alpha$-captures and oxygen-burning produce
silicon-group elements (here, $\silicon[28]$ and $\sulfur[32]$).  In
our calculations, the oxygen-burning
occurs not in the flame itself, but in the hottest parts of the
bounding convection zone.  Fig.~\ref{fig:neflame_structure} shows the
structure of the neon-oxygen-burning flame in our \MESA\ model; note
that the flame thickness is $\sim \unit[10^3]{cm}$ and the flame
velocity is $\sim \unit[0.1]{cm\,s^{-1}}$.

In our \MESA\ calculation we are directly resolving the flame, which
is extremely computationally inefficient.\footnote{The calculation
  shown in Fig.~\ref{fig:neflame} required approximately $3\times10^7$
  timesteps and two wall-clock months on a pair of Intel Xeon E5-2670
  v2 processors.} Future work will benefit from a sub-grid model such
as that used in \citet{Woosley15}, where the flame is not resolved in
the full-star simulation, but tabulated velocities from resolved,
micro-zoned flame calculations are used to propagate a model for the
flame.

\subsection{Silicon Burning and Core Collapse}
\label{sec:siflame}
\label{sec:fe-core}

As neon-oxygen-burning migrates to the center, it leaves behind hot
silicon-group ashes.  Unlike the carbon flame, the neon-oxygen flame does not
fully lift the degeneracy of the material; in model M15,
$\eta_{\mathrm{center}} \equiv \EF / (\kB T) \approx 7$ immediately
after the neon-oxygen flame reaches the center.  Still, some KH contraction
occurs, leading to a series of off-center flashes that consume the
remaining oxygen and neon, converting almost the entire remnant to
silicon-group elements.  (Fig.~\ref{fig:composition} summarizes the
evolution of the composition of the remnant.)  In
Appendix~\ref{sec:critical-mass} we find a critical mass for
off-center silicon ignition of $\approx \unit[1.41]{\Msun}$.  This
suggests that in a narrow range of WD mergers (remnants with masses
$\unit[1.35]{\Msun} \la M \la \unit[1.41]{\Msun}$, accounting for mass
loss), the final merger remnant could be a WD with a silicon-group
composition.

Off-center silicon ignition does occur in model M15, in accordance
with our expectation given its mass, at a mass coordinate of
$M_r \approx \unit[0.75]{\Msun}$; this occurs about 10 years after the
neon-oxygen flame reaches the center.  The conditions at this location (at
the time shown in the bottom panel of Fig.~\ref{fig:composition}) are
$\rho = \unit[2.7 \times 10^8]{\gcc}$ and
$T = \unit[3.6\times10^9]{K}$.  Recently, \citet{Woosley15} discussed
the presence of Si-flashes in the evolution of $\unit[9-11]{\Msun}$
stars.  They found silicon deflagrations beginning in models with CO
core masses of $\approx \unit[1.4]{\Msun}$, with increasingly intense
silcon-burning flashes as the mass increased.  Model M15 ignites
silicon further off-center than any of the models described in
\citet{Woosley15}, so it is difficult to associate model M15 with a
specific model in that work.  However, it does appear that we may be
in a regime where unstable Si-burning plays a role.  Future work will
clarify its importance.


Our calculations are not well-suited to study silicon-burning, so we leave this to future work.  In
particular, we are limited by our use of an $\alpha$-chain network.
For example, this small network does not include electron-capture
reactions that will have occurred during the pre-silicon-burning
evolution, reducing the electron fraction and increasing the abundance
of the more neutron-rich isotopes \silicon[30] and \sulfur[34].  For
now, we provisionally assume that silcon-burning in remnants with
$M \ge \unit[1.41]{\Msun}$ (accounting for mass loss) quiescently
leads to the formation of an Fe-core and that subsequently this
low-mass Fe core will collapse to form a neutron star.  Thus, while we
have revised the evolutionary story for super-Chandrasekhar WD
mergers, we think they are still likely to produce a population of
single, low-mass neutron stars.
More quantitative details about such neutron stars, including the
identification of possible distinguishing characteristics, await models
that have been evolved up to core infall and through core-collapse.

\begin{figure}
  \centering
  \includegraphics[width=\columnwidth]{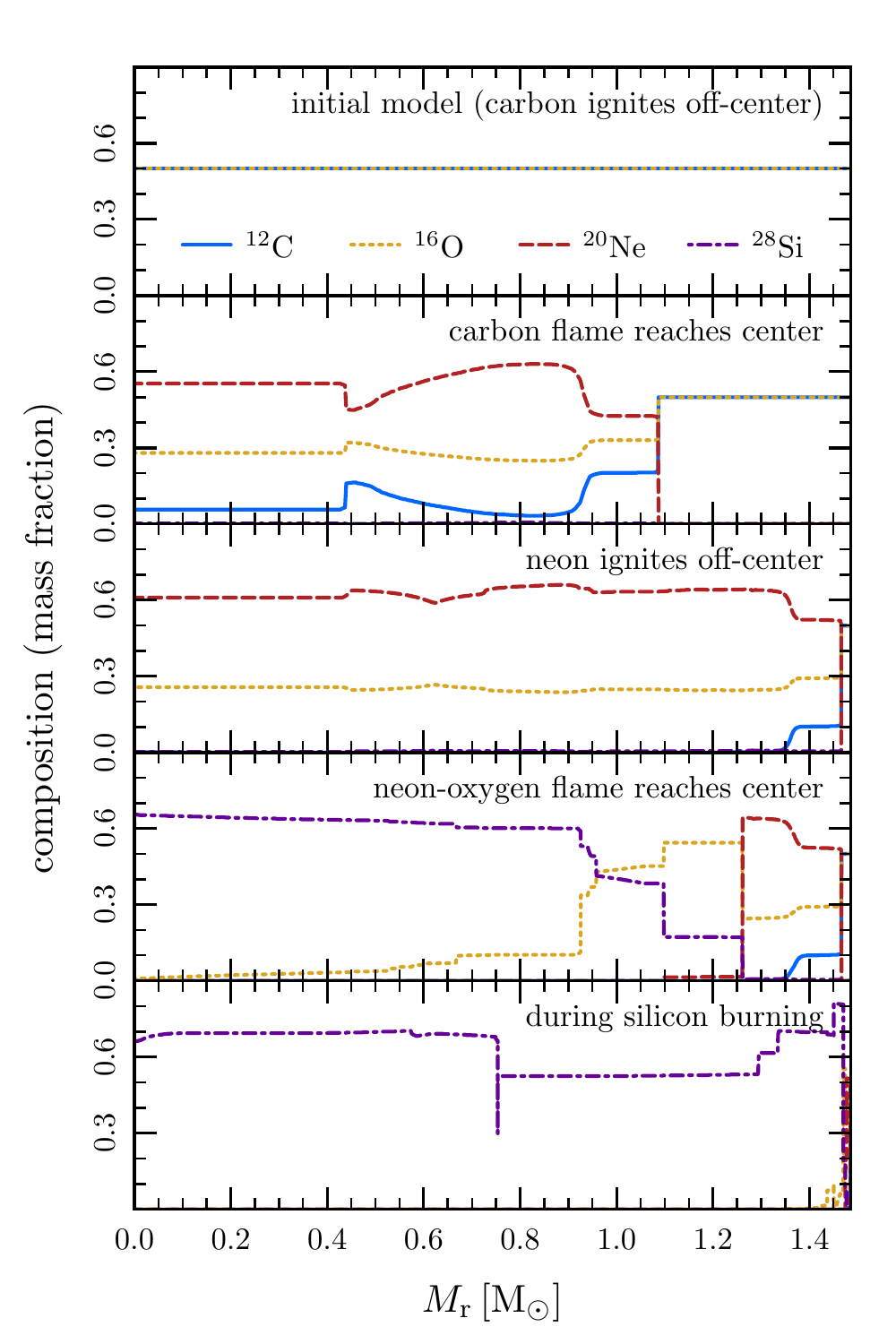}
  \caption{Composition of the merger remnant (model M15 with no mass
    loss) at key times in its evolution. The x-axis shows the
    Lagrangian mass coordinate.  For visual clarity we show only the
    mass fractions of \carbon[12], \oxygen[16], \neon[20], and
    \silicon[28].  The first panel shows the initial model, which is a
    homogeneous carbon-oxygen mixture.  The second panel shows the
    model after the carbon flame has converted the interior to
    oxygen-neon; there is a small amount of residual carbon in the
    region processed by the flame.  The third panel shows the model at
    the time of neon ignition; during the KH contraction phase
    (Section~\ref{sec:kh}) several carbon-burning episodes complete
    the remnant's conversion to oxygen-neon.  The fourth panel shows
    the model after the neon-oxygen flame has converted the core to
    silicon-group isotopes.  The fifth panel shows the model during
    silicon burning; the preceding contraction
    phase has completed the remnant's conversion to silicon-group
    isotopes.}
  \label{fig:composition}
\end{figure}

\section{Observational Properties of the Merger Remnant}
\label{sec:observ-prop}

In this section we describe the observational properties of
super-Chandra WD merger remnants, focusing on the time between the
merger and the final collapse to form a neutron star.  As described in
Sections~\ref{sec:cflame} and~\ref{sec:neflame}, the energy released
by fusion during the carbon and neon-oxygen flames does not reach the
surface.  It is instead lost primarily to thermal neutrino cooling deep in the
stellar interior, at sufficiently high optical depths that the
existence of the flame does not modify the observational properties of
the WD merger remnant.  The latter are instead governed by the heat
released during the merger.

\begin{figure}
  \centering
  \includegraphics[width=\columnwidth]{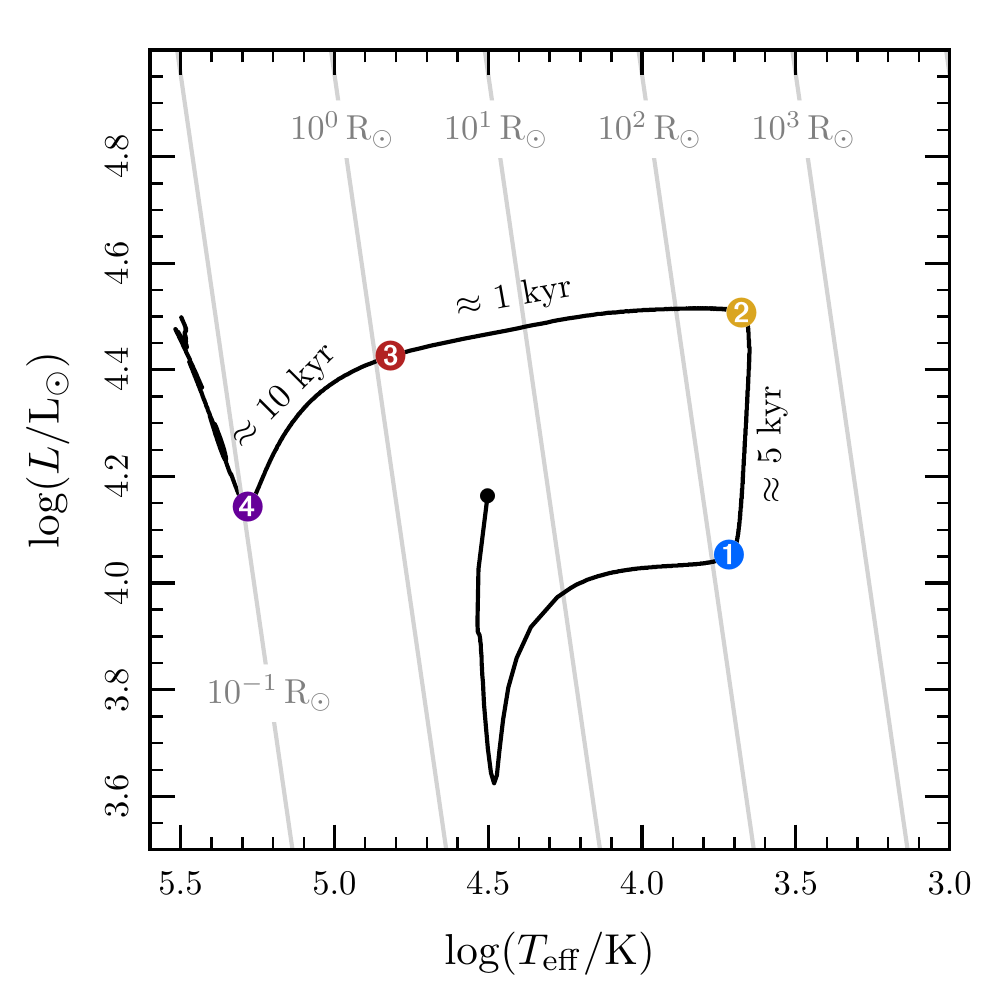}
  \caption{Evolution of model M15 in the HR diagram (model M16 is
    similar).  Grey lines of constant radius are shown in the
    background.  The numbered circles each correspond to one of the
    times indicated in Fig.~\ref{fig:flame-M15}.  The approximate
    elapsed time between adjacent circles is indicated.  The KH
    contraction phase (Section~\ref{sec:kh}) occurs after time 4, with
    the Ne flame phase (Section~\ref{sec:neflame}) corresponding to
    the end of the track.  The total duration of evolution is
    $\approx \unit[25]{kyr}$.  As we discuss in
    Section~\ref{sec:observ-prop}, the presence of a dusty wind around
    these objects may modify their appearance.}
  \label{fig:hr-M15}
\end{figure}

The outer envelope of the remnant responds to the energy deposited
during the merger and begins to radiate away this energy.
Fig.~\ref{fig:hr-M15} shows the location of model M15 in the HR
diagram during our \MESA\ calculation, over the phase in which the
carbon flame is propagating to the center.  The remnant radiates at
the Eddington luminosity for a solar mass object with an effective
temperature $\approx \unit[4000-5000]{K}$ for $\approx \unit[5]{kyr}$,
and then evolves to the blue, spending $\approx \unit[10]{kyr}$ with
$\Teff \gtrsim \unit[10^5]{K}$.

The track shown in Fig.~\ref{fig:hr-M15} for the remnant resembles
that of a star evolving from the AGB to the planetary nebula stage
\citep[e.g.,][]{Kwok93}.  In some ways, the merger has formed an
object similar to the core of an intermediate mass star.  However, the
lack of hydrogen- and helium-burning shells means that thermal pulses
will be absent.  Near the end of the AGB phase, stars are seen to
exhibit extreme ($\Mdot \gtrsim \unit[10^{-4}]{\Msunyr}$) mass loss
rates \citep[e.g.,][]{vanLoon99}.  Winds from these cool, luminous
stars are thought to be launched by a two stage process in which
pulsation-driven shock waves lead to dust formation and then radiation
pressure on the dust accelerates the wind \citep[e.g.,][]{Hofner15}.
The pulsations in AGB stars are driven by H and/or He ionization
\citep{Wood79, Ostlie86}; such pulsations may not be present in pure
carbon-oxygen envelopes.  The structural and compositional differences
between extreme AGB stars and these remnants mean that proximity in
the HR diagram does not itself directly imply analogous mass loss
properties.

The track shown in Fig.~\ref{fig:hr-M15}
does not include the effects of mass loss.  The inclusion of mass loss
has two primary effects: (1) it alters the observational properties of
the remnant, likely obscuring it in an dusty wind and (2) it reduces
the mass of the remnant, potentially influencing the final outcome.
In particular, with $\Mdot \approx \unit[10^{-4}]{\Msunyr}$, over
the duration of the cool giant phase, the remnant would shed
$\Delta M \approx \unit[0.1]{\Msun}$, sufficient to change a model
with total mass $\approx \unit[1.5]{\Msun}$ from super-to-sub
Chandrasekhar (see Section~\ref{sec:kh}).  However, we note that our
model M16 displays similar evolution as M15, but with its higher total
mass $\approx \unit[1.6]{\Msun}$, it can shed
$\Delta M \approx \unit[0.1]{\Msun}$ and remain super-Chandrasekhar.

In Fig.~\ref{fig:hr-M15} the lowest effective temperature
$(\Teff \approx \unit[4000-5000]{K})$ reflects the steep decline in
the opacities at these temperatures---see Section~\ref{sec:opacities}
and Appendix~\ref{sec:kasen-opacities}.\footnote{Our \MESA\ models
  with cool photospheres also have density inversions in the outer
  layers (these are visible in Fig.~\ref{fig:opacity}). In such cases,
  the structure of the 1D models may be different than the solution
  realized in nature, where the energy transport is influenced by
  inherently multi-dimensional effects \citep{Joss73, Jiang15}. This
  introduces an additional uncertainty in our models during this
  phase.}  The R Coronae Borealis stars are giants with similar
effective temperatures $(\unit[4000-7000]{K})$ with He-dominated,
C-enhanced atmospheres \citep[e.g.,][]{Clayton96}.  These objects
exhibit high-amplitude dimming events
(\mbox{$\approx 10\, \mathrm{mag}$} in the optical) which are
understood to be the result of dust formation events outside the
photosphere.  It is possible that the super-Chandrasekhar WD merger
remnants would exhibit similar variability.

Given their cool photospheres between times 1 and 2 in
Fig.~\ref{fig:hr-M15} and their almost pure carbon and oxygen
composition, it seems likely that these remnants would form copious
amounts of dust.  The mass loss rate is then set by the dust formation
rate near the stellar photosphere.  Reprocessing by this dust makes
obscured objects bright infrared (IR) sources, though because the
remnants are hydrogen-free, such an object should not exhibit any
OH-maser emission, as seen in the OH/IR stars \citep{Wilson68}.  The
properties of the dust also depends on the C/O ratio (the number ratio
of carbon and oxygen atoms) and as such, the remnants could manifest
either as C-rich or O-rich objects.  We expect that the surface C/O
ratio will be set largely by the C/O ratio of the disrupted WD.  This
quantity is likely not universal---depending on the mass of the
disrupted WD, for example.

The C/O ratio predicted by stellar models depends on still-uncertain
input physics such as the $^{12}$C($\alpha,\gamma)^{16}\rm{O}$
reaction rate.  The recent study of \citet{Fields16} explored the
effects of this and other reaction rate uncertainties using STARLIB
\citep{Sallaska13}.  For WDs in the mass range
$\unit[0.55-0.65]{\Msun}$, their models using the ``median'' reaction
rates had total C/O ratios in the range $\approx 0.8-1.1$, with this
interval expanding to $\approx 0.7-1.3$ when varying the reaction
rates between their ``low'' and ``high'' values (C.~E.~Fields, private
communication).  Asteroseismic observations of pulsating WDs can
provide observational constraints on the core abundances.
Unfortunately, asteroseismic fits to grids of stellar evolution models
\citep[e.g.,][]{Romero12} have compositions that are ``fixed'' by the
choice of a single set of standard nuclear reaction rates.  Fits using
parameterized composition profiles, such as \citet{BischoffKim14}, do
not suffer from this limitation.  They report core oxygen mass
fractions of a $\unit[0.55]{\Msun}$ WD.  Their two best fit models
(fitting to a grid with a step-size $\Delta X_\mathrm{O} = 0.05$) have
$X_\mathrm{O} = 0.55$ (which is C/O $\approx 1.1$) and
$X_\mathrm{O} = 0.70$ (which is C/O $\approx 0.6$).\footnote{Note that
  treating these core values as estimates of the total fractions
  overestimates the total amount of oxygen, as the models transition
  from an O-dominated interior to a C-dominated exterior.}  We
conclude that it is not yet theoretically or observationally clear
whether we expect the material from the disrupted WD to be C-rich or
O-rich.

If these objects are obscured by a dusty wind, then they may appear in
the same luminosity-color cuts used to identify extreme AGB stars
(see e.g., \citealt{Thompson09b};  approximately 9 of their sources
are coincident with the luminosities of our objects).  In the wind,
$\tau(r) \approx \Mdot\kappa/(4 \pi r v)$, where $\tau$ is the optical
depth, $\kappa$ is the opacity, and $v$ is the wind velocity.  Since
$\tau(\rph) \approx 1$,
\begin{equation}
  \label{eq:rph-remnant}
  \rph \sim \frac{\Mdot \kappa}{4 \pi v} \sim \unit[1.5\times10^{15}]{cm} \left(\frac{\Mdot}{\unit[10^{-4}]{\Msunyr}}\right)
\left(\frac{\kappa}{\unit[10]{cm^2\,g^{-1}}}\right)
\left(\frac{v}{\unit[30]{km\,s^{-1}}}\right)^{-1} ~~~.
\end{equation}
Given the luminosity of our sources, this radius would imply an
effective temperature
\begin{equation}
  \label{eq:Teff-remnant}
  \Teff = \unit[500]{K} 
  \left(
    \frac{L}{10^{4.5} \Lsun}
  \right)^{1/4}
  \left(
    \frac{\rph}{\unit[100]{AU}}
  \right)^{-1/2}
\end{equation}
which is consistent with extreme AGB stars.


We can make a rough estimate of the number of merger remnants in this
cool giant or self-obscured phase at a given time.  The specific rate
of super-Chandrasekhar double WD mergers estimated by
\citet{Badenes12} is
$\unit[1.0^{+1.6}_{-0.6}\times10^{-14}]{yr^{-1}\, \Msun^{-1}}$.  This
implies that the number of sources active in a galaxy is
\begin{equation}
  N_{\mathrm{active}} \approx 1 \times 
  \left(
    \frac{\text{merger rate}}{\unit[10^{-14}]{yr^{-1}\, \Msun^{-1}}}
  \right)
  \left(
    \frac{\text{stellar mass}}{\unit[10^{10}]{\Msun}}
  \right)
  \left(
    \frac{\text{lifetime}}{\unit[10^{4}]{yr}}
  \right)
  \label{eq:active}
\end{equation}
meaning that roughly one of these remnants will be currently active in
M33 and 20 in M31.\footnote{The stellar mass of M33 is
  $\Mstar \approx \unit[3-6 \times 10^9]{\Msun}$ \citep{Corbelli03};
  the stellar mass of M31 is
  $\Mstar \approx \unit[10-15 \times 10^{10}]{\Msun}$ \citep{Tamm12}.}
However, we have no reason to expect sub- and super-Chandrasekhar
models to be observationally distinguishable during this
phase.\footnote{A lower total mass merger will not yet have ignited
  carbon, but recall that the luminous giant phase is powered by
  thermal energy release during the merger, not carbon burning.}
Thus, the number of objects in this phase may be a factor of few
higher, since sub-Chandrasekhar total mass mergers include lower mass
CO WDs closer to the peak of the individual WD mass distribution
\citep{Kepler07}.

Around time 2, in Fig.~\ref{fig:hr-M15}, we expect the extreme mass
loss to end and the dusty envelope to detach from the photosphere.
The evolution from time 2 to time 3, taking $\approx \unit[1]{kyr}$,
is similar to a proto-planetary nebula phase \citep[e.g.,][]{Kwok93}.
This transitions into a planetary nebula phase as the temperature of
the central object rises to $\Teff \gtrsim \unit[10^5]{K}$, making it
a bright extreme ultraviolet / soft X-ray source.  This emission will
begin to ionize material shed during the earlier phase, forming an
ionization nebula.  An evolutionary timescale of
$\sim \unit[10^{4}]{yr}$ suggests a size $\lesssim \unit[0.1]{pc}$ for
this nebula.  The ionized material, which could have a mass
$\sim \unit[0.1]{\Msun}$, will be composed primarily of carbon and
oxygen and should be hydrogen and helium free.  Eq.~\eqref{eq:active}
suggests that there should currently be tens of these nebulae in the
Milky Way.

Depending on the amount of dust formed, the remnant could continue to
remain obscured through these later phases.  The dust surface density
is roughly
\begin{equation}
  \Sigma_{\mathrm{d}} \sim \unit[10^{-5}]{g\,cm^{-2}}
  \left(\frac{f_{\mathrm{dust}}}{1.0}\right)
  \left(\frac{M_{\mathrm{wind}}}{\unit[0.1]{\Msun}}\right)
  \left(\frac{v}{\unit[30]{km\,s^{-1}}}\right)^{-2}
  \left(\frac{\Delta t}{\unit[10^4]{yr}}\right)^{-2}
  \label{eq:dust}
\end{equation}
where $M_{\mathrm{wind}}$ is the total amount of material ejected in
the wind and $f_{\mathrm{dust}}$ is the fraction of this material that
forms dust.  Approximating the cross-section for EUV photons
$(\lambda \approx \unit[10-100]{nm})$ as geometric, the dust opacity
for a grain size $a$ is
$\kappa_{\mathrm{dust}} \approx \unit[10^5]{cm^2\,g^{-1}}
\left(a/\unit[10^{-5}]{cm}\right)^{-1}$, which is roughly appropriate
when $a > \lambda$.  Given Eq.~\eqref{eq:dust}, the dust optical depth
could be $\sim 1$ even $\unit[10^4]{yr}$ after merger.  However, the
hard radiation from the central object may also destroy some of the
dust; the efficiency of this process will depend on both the
composition and grain size \citep[e.g.,][]{Waxman00}.  Whether or not
they remain completely obscured, it seems likely that these objects
would continue to display significant infrared excesses.

During the neon-oxygen flame and later phases of evolution (see
Sections~\ref{sec:kh} and \ref{sec:subsequent-evolution}), the remnant
remains in the vicinity of time 4 (in Fig.~\ref{fig:hr-M15}).  This evolution is relatively rapid $(\approx \unit[5]{kyr})$
and so the object does not yet begin to move down the WD cooling track
if the mass remains above the Chandrasekhar mass.  However, if the
object has shed enough mass that it is below the mass that will lead
to neon ignition, then it will begin to cool.  Continuing the analogy
to existing systems, these objects may then resemble helium-free
versions of GW Vir.  Such objects would likely also exhibit g-mode
pulsations, as these are driven by ionization of carbon and oxygen
\citep{Quirion07}.  As the objects continue to cool, merger remnants
that remain below the Chandrasekhar mass may later appear as carbon
(DQ) WDs \citep{Dufour07, Dufour08}.

\section{Conclusions}
\label{sec:conclusions}

\begin{figure*}
  \centering
  \includegraphics[width=0.85\textwidth]{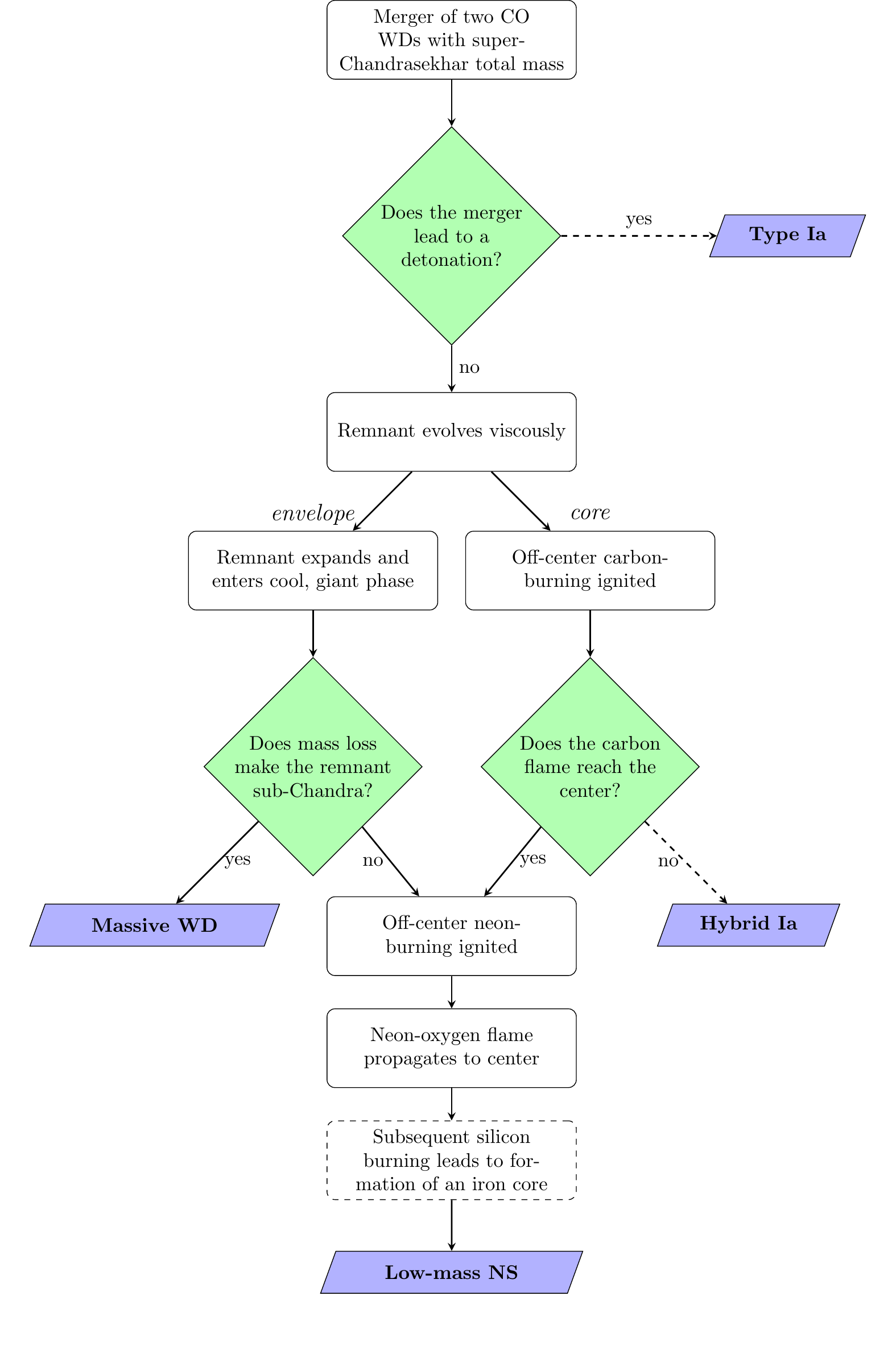}
  \caption{A summary of the final fates of super-Chandrasekhar WD
    merger remnants.  Dashed lines indicate phases not explored in
    this work. Detonations of CO WDs during merger likely requires
    total masses $\ge \unit[2.0]{\Msun}$ \citep{Dan14,Sato15}, so this
    is a rare channel.  Hybrid Ias are unlikely according to
    \citet{Lecoanet16b}.  Thus, most super-Chandrasekhar mergers will
    produce a single massive WD or a low mass NS, with the relative
    rates depending on the uncertain role of mass loss.}
  \label{fig:flowchart}
\end{figure*}

We have presented stellar evolution calculations that follow the
evolution of the remnant of the merger of two CO WDs.  We focused on
systems with a total mass in excess of the Chandrasekhar mass; our
fiducial system is the merger of an 0.6 \Msun\ WD and a 0.9 \Msun\ WD
(model M15 in Table~\ref{tab:models}), but we find similar results for
a system with a total mass of $\unit[1.6]{\Msun}$ (model M16).  Our
calculations use the results of SPH simulations of the merger
\citep{Dan11,Raskin14} to set the initial properties of the remnant.
Its post-merger viscous evolution was then followed as in
\citet{Schwab12} and the results of these simulations form the initial
conditions for our \MESA\ calculations.  Thus, our results do not
apply to the case in which a merger leads to the detonation
of the primary WD, as may occur for particularly massive CO+CO WD mergers \citep{Pakmor12a}.  The flowchart shown in
Fig.~\ref{fig:flowchart} summarizes the potential final fates of these
systems, indicating relevant uncertainties in our models.

Our study has focused on two major questions.  First, what is the
post-merger observational appearance of super-Chandrasekhar WD merger
remnants?  Second, what is the fate of such objects?  The viscous
evolution converts the thick, rotationally-supported disc into a
spherical, thermally-supported envelope.  Our calculations
self-consistently include this material, as opposed to approximating
its effects by imposing an Eddington-limited accretion rate at the
outer boundary.  We can thus better address the emission from WD
merger remnants.  We find a post-merger phase, powered by thermal
energy deposited during the merger, in which the remnant manifests as
an Eddington-luminosity $(L \sim \unit[3\times10^4]{\Lsun})$ source
for $\sim \unit[10^4]{yr}$.  This is consistent, in luminosity and
lifetime, with the signature suggested by \citet{Shen12} who also
constructed related stellar evolution models.  We improve upon these
models by drawing our initial conditions from multi-dimensional
viscous simulations and by incorporating more realistic opacities.

The effective temperature in our models during the giant phase is
$\approx \unit[4000-5000]{K}$.  However, the mass loss properties of
these carbon and oxygen dominated envelopes are uncertain.  It seems
likely that a dusty wind could develop, leading these sources to be
self-obscured and to appear similar to extreme AGB stars.  The mass
loss during the post-merger phase can also influence the later stages
of evolution, leaving an initially super-Chandrasekhar remnant with a
sub-Chandrasekhar mass (see Figs.~\ref{fig:kh-M15-center-mass-loss}
and \ref{fig:flowchart}).  Based on the observed WD merger rates, we
estimate that tens (few) of remnants are presently in this giant phase in
the nearby galaxies M31 (M33).  As the giant phase ends, the central
source evolves, likely producing proto-planetary nebula and planetary
nebula phases in which the surrounding nebula would lack hydrogen or
helium.  These sources might remain dust-obscured throughout,
depending on the amount of dust produced and its properties.
The planetary nebula phase also lasts for $\sim \unit[10^4]{yr}$,
suggesting that there should currently be tens of such sources in
the Milky Way and M31.

In the massive mergers we consider, off-center carbon fusion is
robustly ignited within the remnant.  Most of the fusion energy released is
lost to neutrinos, and that which is not, fails to reach the surface
in the evolutionary time-scale.  Thus, the surface emission is powered
solely by the thermal energy release during the merger.  As found
previously \citep{Saio85, Nomoto85, Saio98}, the carbon-burning
quiescently propagates inward, converting the WD to an oxygen-neon
composition.  However, we follow the evolution of these remnants for
longer than previous calculations and demonstrate that when
carbon-burning reaches the center it lifts the degeneracy of the
remnant.

As the non-degenerate oxygen-neon core undergoes a phase of
neutrino-cooled KH contraction, the remnant can ignite off-center neon
burning.  There is a critical mass for a ``hot'' (i.e.,
non-degenerate) ONe core,
$M_{\mathrm{Ne, hot}} \approx \unit[1.35]{\Msun}$, above which
off-center neon ignition will occur (\citealt{Nomoto84a}; see also
Appendix~\ref{sec:critical-mass}); this is set by the rate of neon
fusion and the rate of thermal neutrino losses.  There is also a
critical mass for electron-capture-initiated central ignition in a
``cold'' (i.e., degenerate) ONe core,
$M_{\mathrm{Ne, cold}} \approx \unit[1.38]{\Msun}$; this is set by the
electron Fermi energy needed to favor electron captures on \neon\
\citep{Schwab15}.  The off-center carbon burning ignited in the merger
remnant leads to the production of a ``hot'' ONe core.  Because
$M_{\mathrm{Ne, hot}} < M_{\mathrm{Ne, cold}}$, we therefore conclude
that lower mass merger remnants ($M < M_{\mathrm{Ne, hot}}$) will form
massive ONe WDs which will then quiescently cool, while higher mass
objects ($M > M_{\mathrm{Ne, hot}}$) will be processed to compositions
beyond neon.  Thus, contrary to standard models, it is difficult to
produce an \textit{oxygen-neon} core with a sufficient mass to undergo
accretion-induced collapse in a WD merger.  We emphasize that our
conclusions apply to WD mergers and do not affect standard single
degenerate AIC scenarios.  In the single degenerate case, an ONe core
below $M_{\mathrm{Ne, hot}}$ forms in an intermediate mass star and
cools.  Subsequently, it accretes material and grows to a mass in
excess of $M_{\mathrm{Ne, cold}}$ at which point it collapses.

One of the major uncertainties in our calculation is the extent of
mass loss during the $\sim \unit[10^4]{yr}$ post-merger thermal evolution.  In addition to our fiducial model, we also evolved a
remnant with a total mass $\approx \unit[1.6]{\Msun}$.  Both models
had similar initial profiles (Fig.~\ref{fig:initial-models}) and
underwent a qualitatively similar course of evolution.  We suspect
that yet higher masses would behave similarly, up to the point where
carbon detonations are likely to occur during the merger.  The results
of both \citet{Dan14} and \citet{Sato15} suggest that the 
threshold for carbon detonations is around a total mass $\ga \unit[2]{\Msun}$ for CO+CO WD mergers.
Since this threshold is significantly super-Chandrasekhar, this
suggests that even in the presence of significant mass loss
($\approx \unit[0.2]{\Msun}$) both the massive WD and the low mass NS
fates in Fig.~\ref{fig:flowchart} are realized. However, the exact
mapping of merger masses to outcome is uncertain.

If the remnant mass remains super-Chandrasekhar and experiences
off-center neon ignition, we expect that it will ultimately collapse
to form a neutron star.  The neon-oxygen flame propagates to the center
converting the remnant to silicon-group elements
(Figs.~\ref{fig:neflame} and \ref{fig:neflame_structure}).
Computational limitations prevent us from continuing our calculations
much beyond the point at which the neon-oxygen flame reaches the center.
However, we then argue, based on the work of \citet{Jones13, Jones14}
and \citet{Woosley15} that Si-burning leads to the formation of an
iron core, which will then collapse.  Future work in this area will be
important in understanding the role of violent Si-burning and
examining the structure of the remnant at the time of core infall.
The low mass core, with steep density gradients near its edge, means
that this progenitor will likely produce a neutrino-driven explosion a
short time after core bounce, as is seen in core-collapse calculations
involving ONe cores \citep{Kitaura06, Janka08}.  A shorter time
between core bounce and explosion also implies that there is less time
for instabilities to grow, leading to less asymmetry and lower neutron
star kicks \citep[e.g.,][]{Wongwathanarat13}.  With pre-collapse
structures in-hand, future explosion calculations can determine the
amount of radioactive material produced and thus the brightness of the
accompanying supernova---simulations of ONe cores find only
$\unit[10^{-3}-10^{-2}]{\Msun}$ of \nickel[56] \citep{Dessart06,
  Kitaura06}---but the amount of \nickel[56] synthesized in remnants
that remain significantly super-Chandrasekhar at the time of collapse
remains to be determined.

Our calculations assume that the bounding convection zone does not
induce significant mixing in either the carbon flame or the neon
flame.  The results of \citet{Lecoanet16b}, which we invoked to
justify this assumption in the case of the carbon flame do not
strictly apply to the neon-oxygen flame.  The neon-oxygen flame self-crossing time
(which is equivalent to the burning time-scale) is much shorter,
$\sim \unit[10^4]{s}$, on the order of the convective turnover time in
the zone behind the flame.  This invalidates the assumption made by
\citet{Lecoanet16b} that the flame is effectively stationary over a
convective turnover time.  Na\"{i}vely, however, it seems likely that
fewer convective turnover times make it even less likely that
convection will disrupt the flame, as convective perturbations will
not be able to accumulate because the flame moves away.

Because the duration of the giant phase is shorter than the time for
the carbon flame to reach the center, our results indicate that at the
time of neutron star formation the remnant is likely in a compact
configuration (Fig.~\ref{fig:hr-M15}), implying minimal signatures due
to interaction or shock breakout.  However, if violent Si-flashes unbind some mass, as \citet{Woosley15} found in some of their analogous intermediate mass star models, this could produce strong circumstellar interaction in the resulting supernovae.  At late times after the supernova
explosion ($\sim \unit[10]{yr}$), the ejecta from the explosion (with
$v \sim \unit[10^4]{km\,s^{-1}}$) will catch up to the wind generated
during the giant phase (with $v \sim \unit[10]{km\,s^{-1}}$); however
this interaction is radiatively inefficient.

Future work should continue to explore the variety of final outcomes
shown in Fig.~\ref{fig:flowchart} and work to clarify further
differences between neutron stars formed via single-degenerate
accretion-induced collapse and those formed in WD mergers.

\section*{Acknowledgments}
We thank Lars Bildsten, Jared Brooks, Rob Farmer, Jason Ferguson, Ken
Shen, and Frank Timmes for helpful discussions.  We thank Marius Dan
and Cody Raskin for providing the results of their SPH simulations as
part of previous work.  We thank Ken'ichi Nomoto, Todd Thompson, and
Stan Woosley for useful conversations following the presentation of
these results in preliminary form.  We thank an anonymous referee for
comments that led to improvements in the manuscript.  We acknowledge
stimulating workshops at Sky House and Palomar Observatory where these
ideas germinated. JS is supported by the NSF Graduate Research
Fellowship Program under grant DGE-1106400 and by NSF grant
AST-1205732. EQ is supported in part by a Simons Investigator award
from the Simons Foundation and the David and Lucile Packard
Foundation.  This research is funded in part by the Gordon and Betty
Moore Foundation through Grant GBMF5076.  DK was supported in part by
a Department of Energy Office of Nuclear Physics Early Career Award,
and by the Director, Office of Energy Research, Office of High Energy
and Nuclear Physics, Divisions of Nuclear Physics, of the
U.S. Department of Energy under Contract No.  DE-AC02-05CH11231.  This
research used the SAVIO computational cluster resource provided by the
Berkeley Research Computing program at the University of California
Berkeley (supported by the UC Chancellor, the UC Berkeley Vice
Chancellor of Research, and the Office of the CIO).  This research has
made use of NASA's Astrophysics Data System and GNU Parallel
\citep{Tange11}.



\bibliographystyle{mnras}
\bibliography{COMergers}


\appendix

\section{Importing Models into \MESA}
\label{sec:mapping-models}

Mapping results from one simulation code into another can be a
difficult process.  Differences in the input microphysics, the
equations being solved, and the numerical solution techniques mean
that output of one code is rarely immediately suitable for input into
another code \citep[e.g.,][]{Zingale02}.  Each approach is problem-specific: one must identify
the important parts of the solution and design a mapping that
preserves them.  Our goal is to use the output of a 2D Eulerian code
(ZEUS-MP2, which solves the viscous fluid equations) as input to a 1D
Lagrangian code (\MESA, which solves the stellar structure
equations).



At the end of our hydrodynamic simulations of the viscous phase of
evolution, most of the WD merger remnant is slowly rotating and in
hydrostatic equilibrium and thus quite spherical \citep[e.g., fig.~5
in][]{Schwab12}.  We make the choice to study the non-rotating case
and thus do not initialize or evolve a rotational velocity variable.
Under the assumption of spherical hydrostatic equilibrium, the
complete structure of an object can be specified by its specific
entropy, $s(M_r)$, and composition, $X_i(M_r)$, as a function of
Lagrangian mass, $M_r$.  Therefore, the first step is to spherically
average our multi-D simulations and calculate these quantities.

We perform volume averages\footnote{We write these averages as
  integrals, but because we are operating on the output of a
  grid-based code, these integral quantities represent the
  appropriate discretized sums.},
\begin{align}
  \rho(r) & = \frac{1}{2}\int_{0}^{\pi} d\theta \sin(\theta) \rho(r,\theta)\\
  e(r) & = \frac{1}{2}\int_{0}^{\pi} d\theta \sin(\theta) e(r,\theta)\\
   \rho(r) X_i(r) & = \frac{1}{2}\int_{0}^{\pi} d\theta \sin(\theta) \rho(r,\theta) X_i(r,\theta)
\end{align}
so that the appropriate quantities (e.g. mass, energy) are conserved.
Assuming full ionization, it is simple to calculate $\bar{A}$ (the
average mass per ion) and $\bar{Z}$ (the average charge per ion) from
the mass fractions $X_i$.  Given $\rho$, $e$, $\bar{A}$, and
$\bar{Z}$, we use the Helmholtz \citep{Timmes00b} equation of state,
which is used by both our ZEUS-MP2 simulations and \MESA\ (for
$\rho \ga \unit[3\times10^2]{g\,cm^{-3}}$), to calculate the specific
entropy $s(r)$.  The Lagrangian mass is
\begin{equation}
  M_r = \int_{0}^{r} dr' 4 \pi r'^2 \rho(r')
\end{equation}
and we record a 1D approximation to our 2D simulation consisting of the
values of $M_r$, $s(r)$, and $X_i(r)$ for each radial grid point in
the computational domain.

We want to create a \MESA\ model which matches this profile.  Instead of
trying to create a \MESA\ model file and then reading it in, we begin with
a model unlike what we want, but slowly reshape it into our desired
profile.  The steps in this procedure were arrived at by
trial-and-error.  There is nothing inherently correct about many of
the particulars of this approach; we demonstrate at the end that this
stellar engineering gives us the desired result.

First, we create a pre-main sequence model with the mass of the
remnant.  We evolve this model, with nuclear reactions turned off,
until the central density is equal to $\unit[10^3]{g\,cm^{-3}}$.  Then
we resume the evolution, using the built-in capability of \MESA\ to
relax our model to a specified composition.  With the composition
relaxation complete, we evolve this model, with nuclear reactions and
mixing turned off, until the central density reaches
$\unit[10^5]{g\,cm^{-3}}$.

Again, we resume the evolution, this time making use of a custom
routine in \texttt{run\_star\_extras.f} which relaxes the model to the
desired thermodynamic profile.  We take advantage of the
\texttt{other\_energy} routine, which allows us to add an additional
term to the energy equation, and set
 \begin{equation}
   Q_\mathrm{extra, k} = -\frac{c_{v,k} T_k}{t_o} \alpha_k
 \end{equation}
where $c_v$ is the specific heat at constant volume and $T$ is the
temperature.  The subscripts $k$ indicate that these quantities are
evaluated on a per-cell basis.  The values of $\alpha_k$ were
chosen to drive the model towards the desired profile.  We chose a
heating/cooling time-scale, $t_0$, which is short compared to the
thermal time of the star and is comparable to the total duration of
our viscous evolution calculations, typically $t_0 = \unit[10^3]{s}$.

To determine $\alpha_k$, we read in the spherically-averaged entropy profile from ZEUS and use built-in \MESA\
interpolation routines to sample it such that we have a target entropy
function, $\tilde{s}(q)$, where $q = M_r / M_\mathrm{tot}$.  In the
inner part of the model, which corresponds to the cold, undisturbed
primary WD, we will ignore the entropy profile and instead relax the
profile to an isothermal one.  This approach requires three additional
parameters: the isothermal temperature, $\tilde{T}$, the region which
should be isothermal, $\tilde{q}$, and the width of a region in which
we blend between these, $\Delta$.  At each point in the \MESA\ model, we
evaluate
\begin{align}
  \delta_1 & = 1 - \tilde{s}(q_k)/s_k\\
  \delta_2 & = 1 - \tilde{T}/T_k \\
  \delta_k & = f_k \delta_1 + (1-f_k) \delta_2
\end{align}
where 
\begin{equation}
  f_k = \frac{1}{2} \left[\tanh\left(\frac{q_k - \tilde{q}}{\Delta}\right) +1 \right]
\end{equation}
From these quantities, we locally calculate
\begin{equation}
  \alpha_k = \delta_k \tanh(|\delta_k|)  
\end{equation}
and in order to check whether the profile matches, calculate a global
quantity
\begin{equation}
  \epsilon = \sum_{k = 1}^{n_z} |\delta_k| dq_k
\end{equation}
We consider our relaxation complete when
$\epsilon < 3 \times 10^{-4}$.  Empirically, this approach works
quickly and robustly.

\begin{figure}
  \centering
  \includegraphics[width=0.48\textwidth]{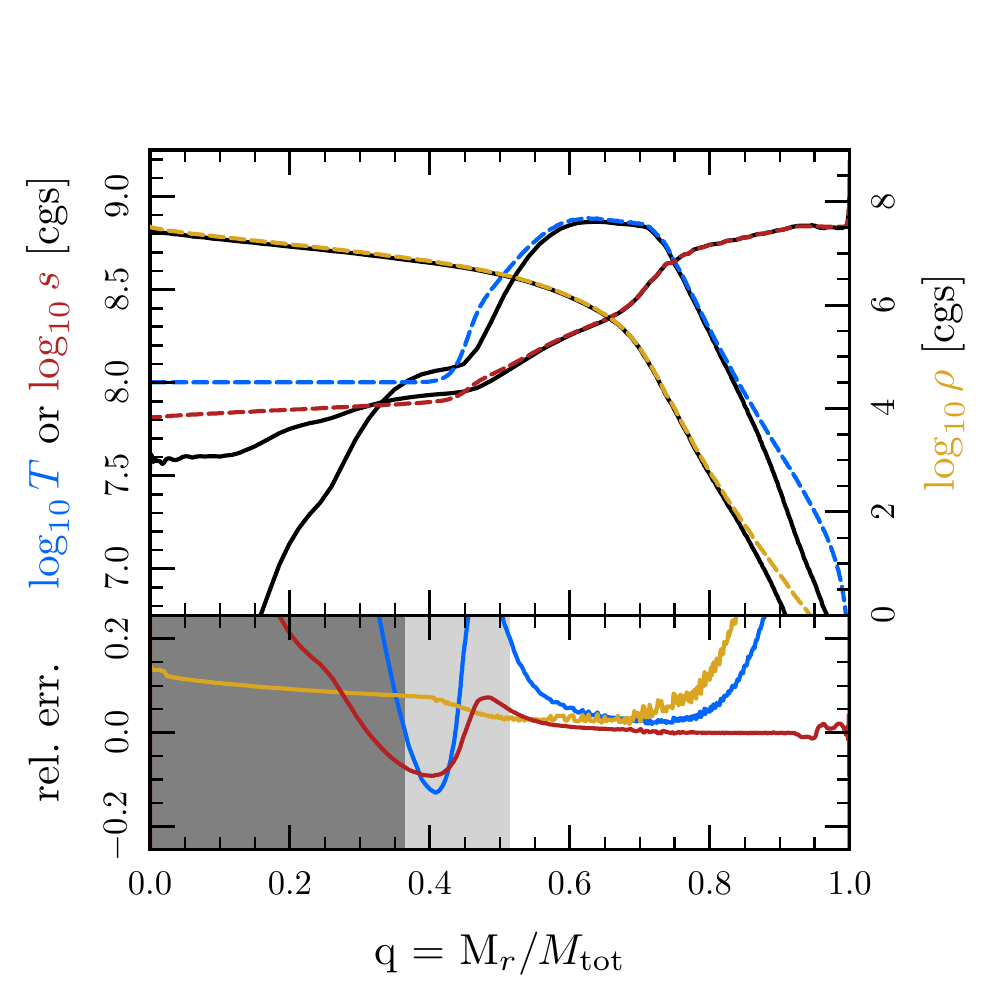}
  \caption{A comparison of the 1D-averaged ZEUS-MP2 model and the
    initial \MESA\ model for our model M15. The top panel overplots the
    1D-averaged ZEUS-MP2 models (black solid lines) and the initial
    \MESA\ model (colored dashed lines).  Shown are the temperature
    (blue; in $\unit{K}$), specific entropy (red; in
    $\unit{erg\,s^{-1}\,g^{-1}}$), and density (yellow; in $\unit{g\,cm^{-3}}$).
    The bottom panel shows the relative error in the three
    thermodynamic quantities.  The dark shaded region indicates where
    the model was relaxed to an isothermal profile and the unshaded
    region shows where the model was relaxed to a given entropy profile (see Appendix~\ref{sec:mapping-models}).
    The lighter shaded region shows the region which blends between
    the two.}
  \label{fig:mapping}
\end{figure}

Fig.~\ref{fig:mapping} compares the 1D-averaged ZEUS-MP2 profiles and
the initial \MESA\ profiles for our model M15.  The density, entropy,
and temperature are in good agreement for $0.5 \la q \la 0.9$.  This
is the critical region, because it is where carbon ignition will occur
and where the thermal energy that will power the giant phase is
located.  In the deep interior $(q \la 0.5)$, our assumed
isothermality means the temperature and entropy of the material has
been altered, but because the material is degenerate, this has little
effect on the structure (i.e., the density agrees well).  In the outer
regions $(q \ga 0.9)$, modest differences are introduced by the fact that the
material was not spherical, but is now assumed to be.

\section{Opacities}
\label{sec:kasen-opacities}

As discussed in Section~\ref{sec:opacities}, we generate and use a set
of low temperature opacities for material with a carbon-oxygen
composition.  The code calculates Rosseland mean opacities considering
the bound-bound, bound-free, free-free, and electron-scattering
contributions.  It assumes local thermodynamic equilibrium and all
photoionization cross-sections are assumed to be hydrogenic.  The
input atomic data is that compiled as part of CMFGEN
\citep{Hillier11}.  We calculated and used a table for a single
composition: $dX_\mathrm{C} = 0.49$, $dX_\mathrm{O} = 0.49$, $Z = 0.02$,
where the relative metal abundances are drawn from \citet{Grevesse98}.

\begin{figure*}
  \centering
  \includegraphics[width=0.48\textwidth]{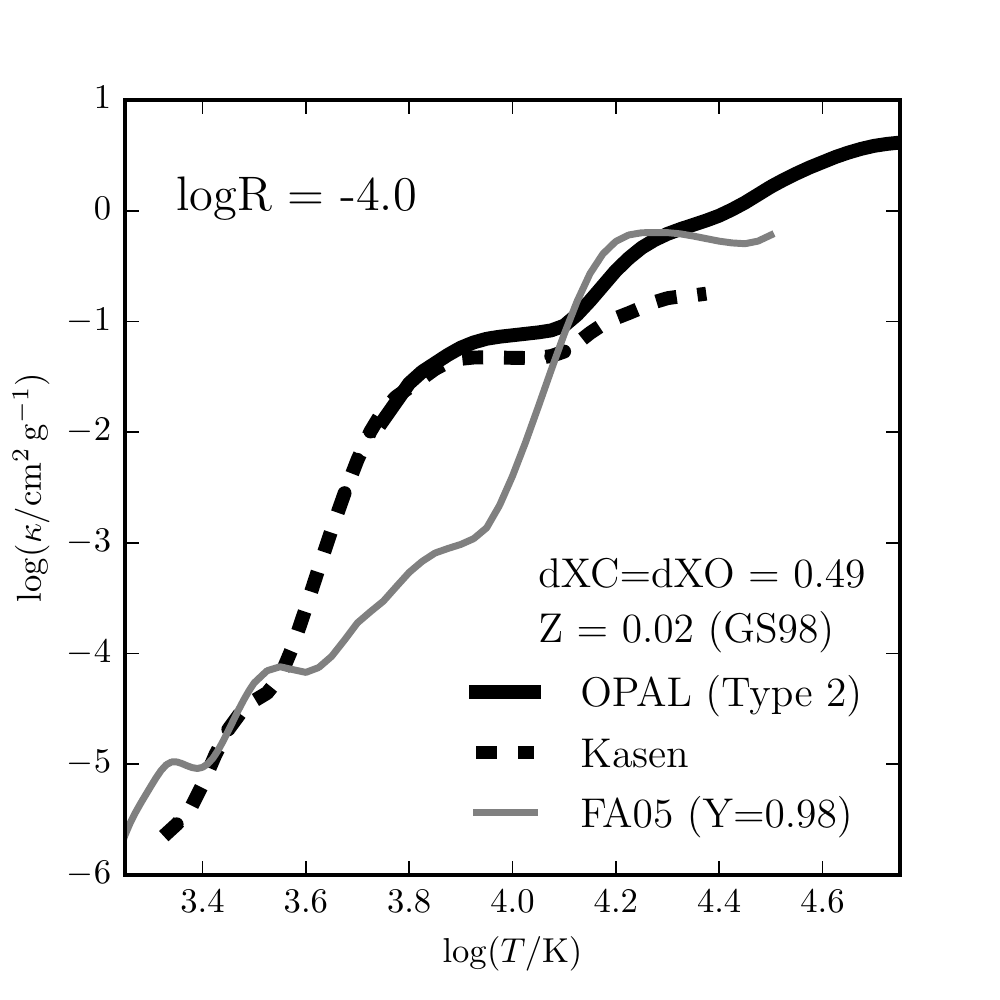}
  \hfill
  \includegraphics[width=0.48\textwidth]{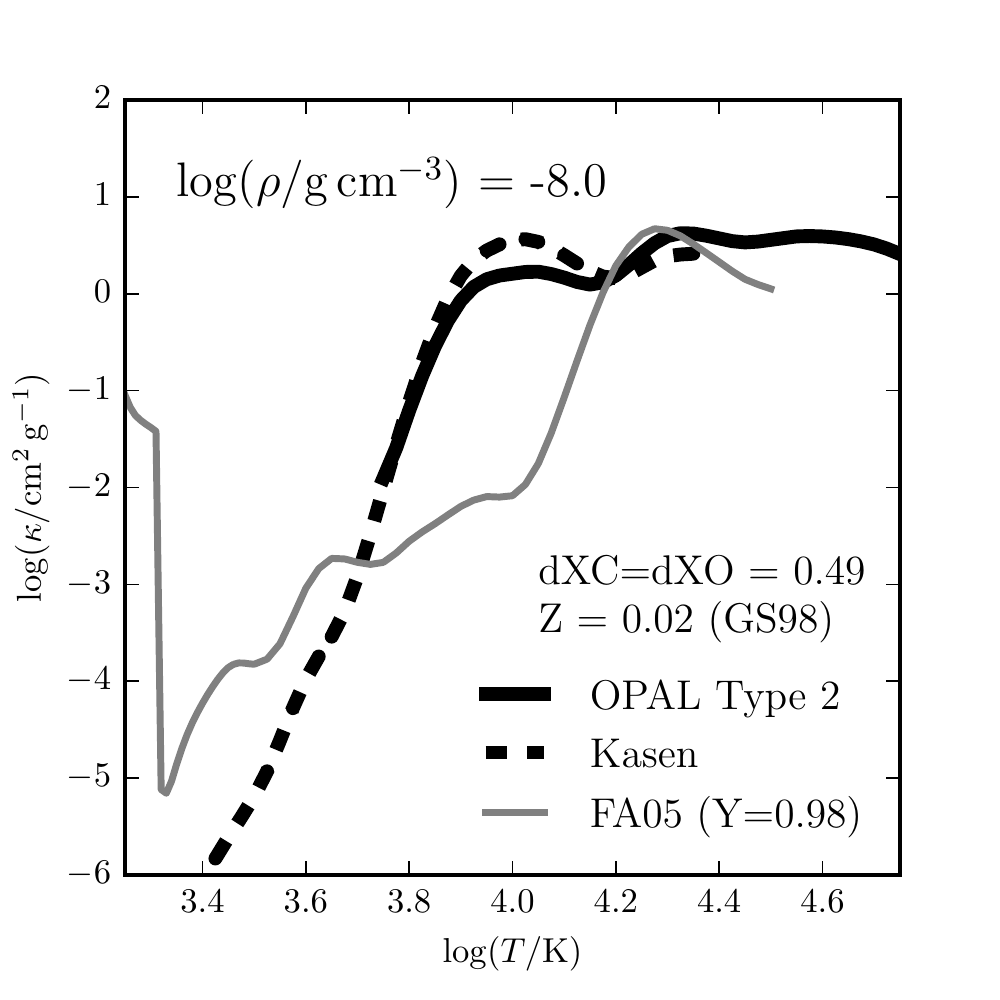}
  \caption{The solid black line (labeled OPAL Type 2) is the carbon-
    and oxygen-enhanced radiative opacities from \citet{Iglesias96}.
    The calculations presented in this paper use low-temperature
    opacities calculated with a code developed by one of us (Kasen;
    dashed black line).  The thin grey line (labeled FA05) is the
    low-temperature opacities from \citet{Ferguson05}; these are the
    opacities that \MESA\ would use if we did not provide our own.
    The difference between the OPAL/Kasen and FA05 curves are because
    \MESA\ has assumed a helium-dominated composition (see text
    for discussion).  In all cases, the relative metal abundances are
    drawn from \citet{Grevesse98}.  In our \MESA\ calculations, we
    blend the OPAL and Kasen opacities between $\logT = $ 4.1 and 4.2
    (see Fig.~\ref{fig:opacity}).  The left panel shows the opacity as
    a function of temperature at fixed
    $\log R \equiv \log \rho - 3 \log T + 18$ (cgs); the right panel
    shows it at fixed $\rho$. }
  \label{fig:kasen-opacities}
\end{figure*}

\begin{figure}
  \centering
  \includegraphics[width=\columnwidth]{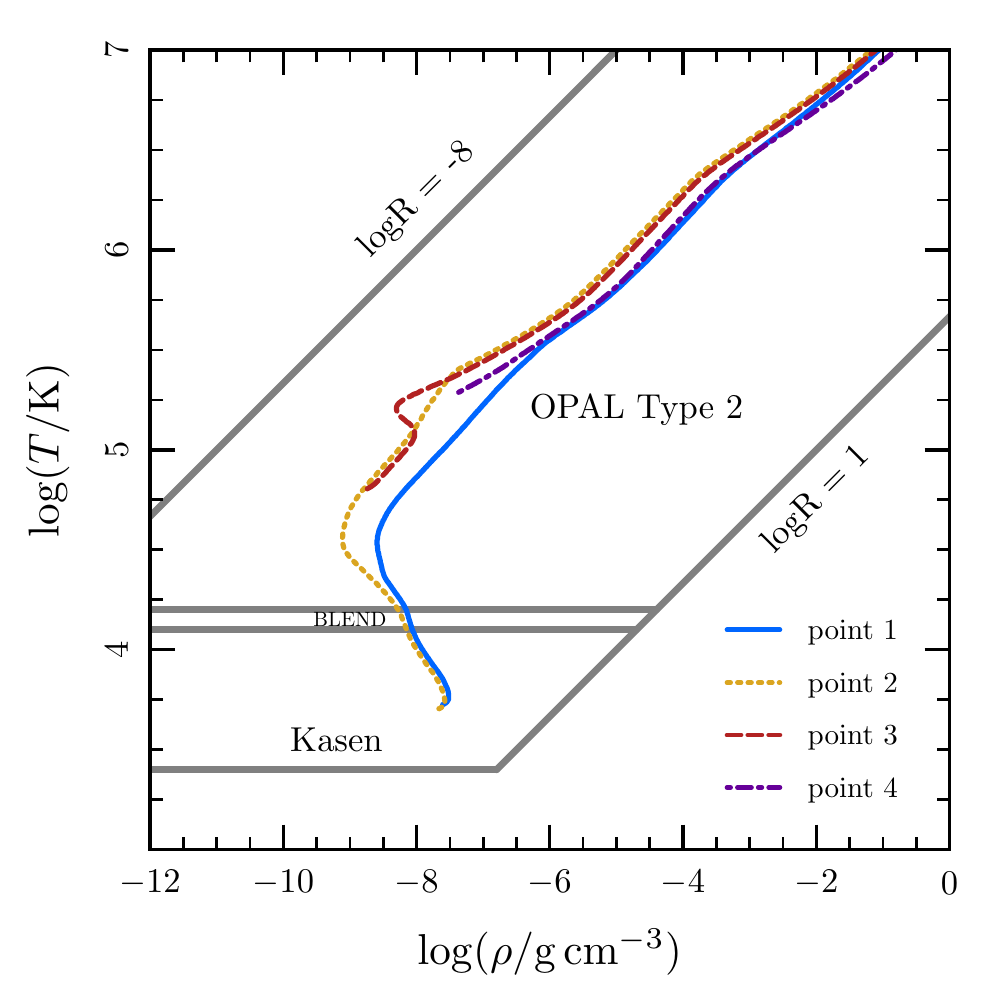}
  \caption{Coverage of temperature-density space by the opacity tables
    used in our \MESA\ calculation.  Solid grey lines show the
    boundaries of the two types of opacity tables: OPAL Type 2
    \citep{Iglesias96} and Kasen (Appendix~\ref{sec:kasen-opacities}).
    ``Blend'' indicates a smooth blend between these two
    values. Recall $\log R \equiv \log \rho - 3 \log T + 18$ (cgs).  The
    other lines show the temperature-density profiles of our model at
    the set of times also marked in Figs.~\ref{fig:flame-M15},
    \ref{fig:center-M15}, and \ref{fig:hr-M15}.  We observe density
    inversions at both the Fe opacity bump $(\logT \approx 5.3)$ and
    in bumps associated with carbon and oxygen.}
  \label{fig:opacity}
\end{figure}

Fig.~\ref{fig:kasen-opacities} shows examples of these opacities near
conditions where they are applied in our models.  We also show the
higher temperature opacities that we use (OPAL Type 2) and the low
temperature opacity that \MESA\ would use if we did not override this
choice via the \texttt{other\_kap} routines.  In our calculation, we
blend the OPAL and Kasen opacities between $\logT = 4.1$ and $4.2$.
Fig.~\ref{fig:opacity} shows the regions of temperature-density space
covered by each of the opacity tables used in our \MESA\ calculation.
Additionally, the temperature-density profiles of model M15 at the
same 4 times indicated in Fig.~\ref{fig:hr-M15} are shown.

The OPAL Type 2 tables are compiled as functions of $X$, $Y$, and $Z$
(H, He, and metals) as well as $dXC$ and $dXO$ (carbon and oxygen
enhancements).  At this time, low-temperature tables that incorporate
the effects of C and O enhancement are not included in \MESA.
Therefore, when transitioning to low temperature opacities, \MESA\ is
also transitioning into a region in which the opacities are tabulated
only as a function of $X$, $Y$, and $Z$.  Thus, the compositions
assumed in the calculation of the low and high temperature opacities
are necessarily different.  The carbon and oxygen composition of our
WD model nominally corresponds to $Z=1$, but since the assumed
abundance distribution within $Z$ is based on solar abundances, a
choice of $Z=1$ effectively assumes the material is dominated by \nitrogen\ and
\iron.  One of the ways \MESA\ can handle this is to assume a
user-specified value of $Z$ (\texttt{Zbase}), use the value of $X$,
and put the rest of the abundance in $Y$.  The line for the low
temperature FA05 opacities in Fig.~\ref{fig:kasen-opacities} makes
this choice, assuming $Z=0.02$ and thus $X=0$, $Y=0.98$.  In addition
to being physically inconsistent, these choices are numerically
unsatisfactory because they result in large change in the opacity and
its derivatives over a small temperature range.  Our use of the Kasen
opacities avoids both of these issues.

The opacities used in the \MESA\ calculations do not include the
effects of molecules.  If we were to include these effects, the
opacity at $\logT \lesssim 3.5$ would be extremely sensitive to the C/O
ratio of the material \citep[e.g.,][]{Ferguson08}.  This quantity was
not, however, self-consistently determined in our initial WD models, which
have equal mass fractions of \carbon[12] and \oxygen[16].


\section{Critical Masses}
\label{sec:critical-mass}

For a contracting stellar model of a given composition, there is a
critical mass above which the temperature will become sufficiently
high to ignite nuclear fusion.  This is familiar for typical
hydrogen-rich compositions, where this critical mass marks the
boundary between a brown dwarf and a low mass star.  \citet{Nomoto84b}
calculated this critical mass for pure neon stars and used this simple
model to gain insight into the evolution of contracting ONe cores.
For pure neon models, the higher density and temperatures mean the KH
contraction is driven by neutrino cooling from the interior as opposed
to photon cooling from the surface.  The density dependence of the
neutrino cooling rates lead to the formation of an off-center
temperature peak, and when ignition does occur, it occurs off center.

In the spirit of \citet{Nomoto84b}, we run a series of \MESA\ models
with masses in the range $1.30\,\Msun$ to $1.44\,\Msun$, using a
stride of $0.005\,\Msun$, and with neon, oxygen-neon, oxygen, and
silicon compositions.  We create a pre-main sequence model with the
desired mass and then relax the model to the desired composition.
With nuclear reactions off, we evolve the model until $\logRhoc = 4$.
At this point, we load the model using the \texttt{approx21} nuclear
network.  We then evolve each model until either (1) the rate of
energy release from nuclear reactions exceeds the rate of energy loss
from neutrinos anywhere in the star and subsequently exceeds
$\unit[10^7]{erg\,s^{-1}\,g^{-1}}$, in which case we classify the
model as having ignited or (2) the peak temperature reaches a maximum
value and subsequently declines, in which case we know the model will
never ignite.

\begin{figure}
  \centering
  \includegraphics[width=\columnwidth]{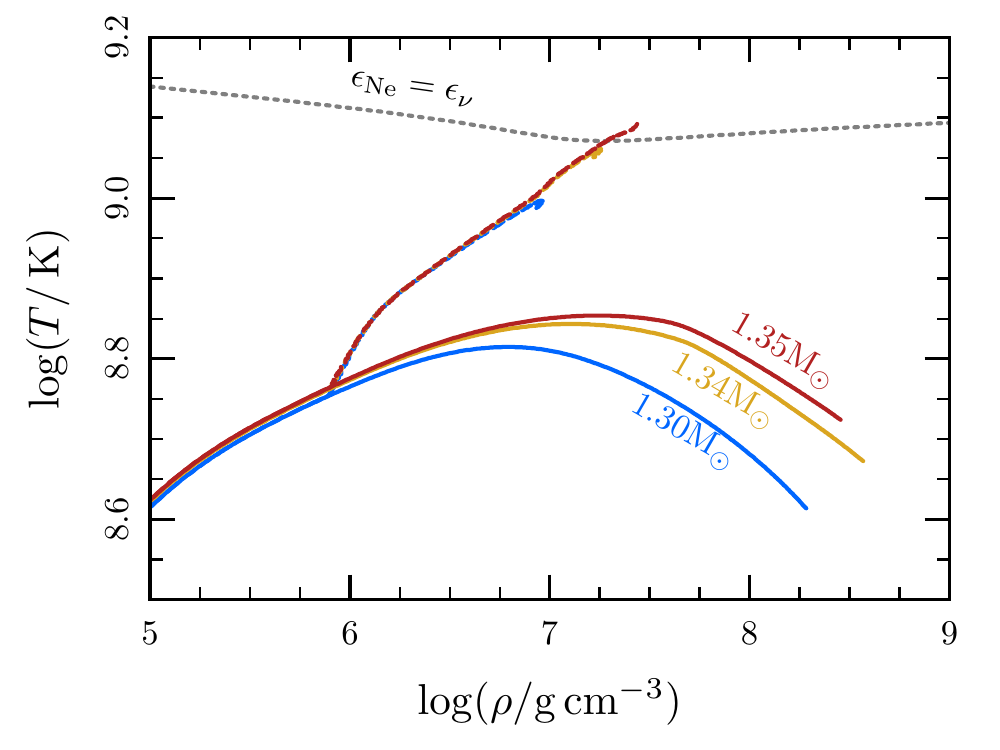}
  \caption{The evolution of temperature and density at the center
    (solid lines) and at the temperature peak (dashed lines) during
    the KH contraction of pure neon models.  The dotted line shows
    approximately where the rate of energy generation from neon burning
    is equal to the energy loss rate from thermal neutrinos.  The
    energy generation rate for neon-burning is that given in
    \citet{Woosley02}. This figure is a reproduction using \MESA\ of
    the results presented in fig.~1 of \citet{Nomoto84b}.  We find a
    slightly lower critical mass for neon ignition of 1.35 \Msun.}
  \label{fig:nomoto84}
\end{figure}

In Fig.~\ref{fig:nomoto84}, we show the evolution of three of our pure
neon models.  We find the lowest mass model that ignites has
$M = 1.35\,\Msun$.  This is slightly lower than the value of
$1.37\,\Msun$ reported by \citet{Nomoto84b}.  We speculate that this
minor difference is due to differences in the input microphysics.
Curves analogous to these, but for different masses, are shown as the
grey lines in Figs.~\ref{fig:kh-M15-center} and
\ref{fig:kh-M15-center-mass-loss}. The qualitative agreement between
our WD merger remnant models and these simple homogeneous models demonstrates
that the off-center neon ignition found in our WD merger remnants is a simple
consequence of their mass.

\begin{figure}
  \centering
  \includegraphics[width=\columnwidth]{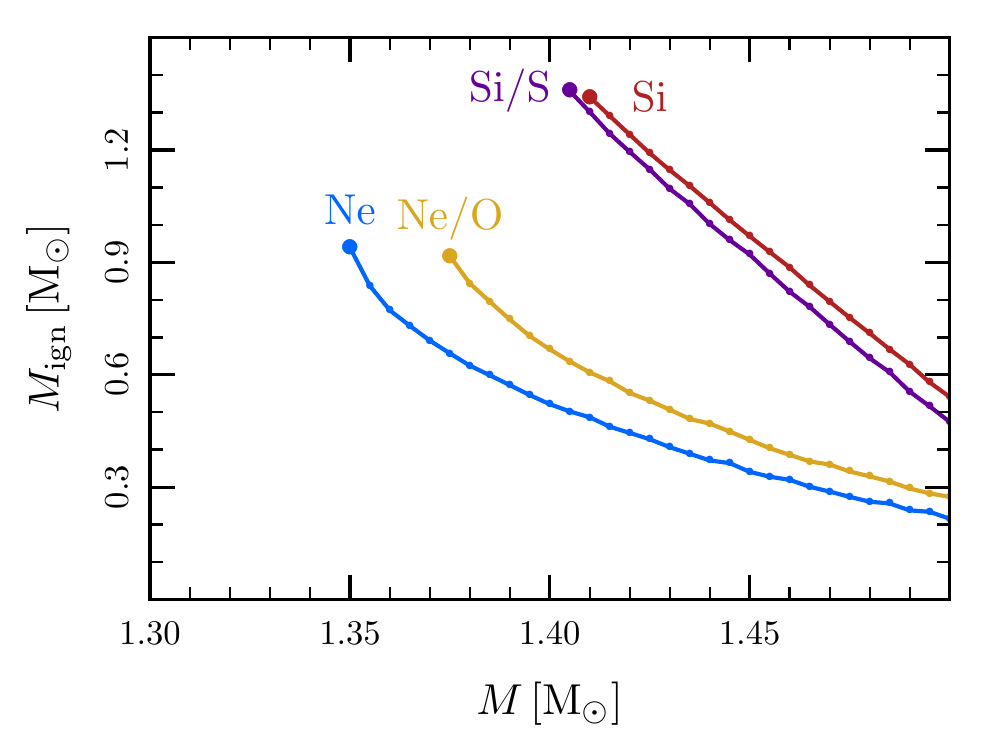}
  \caption{The mass coordinate of off-center ignition for different
    assumed compositions: pure \neon[20] (Ne), an equal mixture by
    mass of \neon[20] and \oxygen[16] (Ne/O), pure \silicon[28] (Si),
    and an equal mixture by mass of \silicon[28] and \sulfur[32]
    (Si/S).  Each model is indicated with a dot; the lowest mass model
    that ignited is indicated with a large dot.  The dots are
    connected by lines to guide the eye.}
  \label{fig:ignition-coordinate}
\end{figure}

The mass coordinate at which ignition occurs is shown in
Fig.~\ref{fig:ignition-coordinate}.  We show 4 different compositions:
pure \neon[20] (Ne), an equal mixture by mass of \neon[20] and
\oxygen[16] (Ne/O), pure \silicon[28] (Si), and an equal mixture by
mass of \silicon[28] and \sulfur[32] (Si/S).  The large dots indicate
the lowest mass models in which off-center ignition was observed;
these are the critical ignition masses quoted in the text.  Caveats
apply to our treatment of silicon-burning: we have used a small
$\alpha$-network and neglected neutron-rich isotopes such as
\silicon[30] and \sulfur[34].  However, since the same caveats apply
in our calculation of the remnant evolution, these simple models are
well-suited for use in interpreting our results.

\bsp	
\label{lastpage}
\end{document}